\documentclass[a4paper]{aa}%
\usepackage{graphicx}
\usepackage[round]{natbib}
\usepackage{fancyhdr}
\usepackage{amsmath}
\usepackage{subfig}

\newcommand{\macc}{$\dot{M}_{acc}$}
\newcommand{\nep}{[Ne\,{\sc ii}]}

\newcommand{\cp}{[C\,{\sc ii}]}
\newcommand{\arp}{[Ar\,{\sc ii}]}

\newcommand{\on}{[O\,{\sc i}]}

\newcommand{\lx}{$L_{\rm X}$}
\newcommand{\luv}{$L_{\rm FUV}$}
\newcommand{\mic}{$\rm \mu m$}
\newcommand{\apo}{$a_{\rm pow}$}
\newcommand{\ami}{$a_{\rm min}$} 
\newcommand{\eps}{$\epsilon$}
\newcommand{\rin}{$R_{\rm in}$}
\newcommand{\rou}{$R_{\rm out}$}
\newcommand{\tave}{\left< T\right>}

\begin{document}
\title{FUV and X-ray irradiated protoplanetary disks: a grid of models}
\subtitle{II - Gas diagnostic line emission}
\author{G.~Aresu\inst{1}
  \and R.~Meijerink\inst{1,2}
  \and I.~Kamp\inst{1}
  \and M.~Spaans\inst{1}
  \and W.-F.~Thi\inst{3}
  \and P.~Woitke\inst{4}}
\institute{Kapteyn Astronomical Institute, Postbus 800, 9700 AV Groningen, The Netherlands
  \and Leiden Observatory, Leiden University, P.O. Box 9513, NL-2300 RA, Leiden, The Netherlands
  \and UJF-Grenoble, CNRS-INSU, Institute de Plan\`etologie et d'Astrophysique (IPAG) UMR 5274
  \and SUPA, School of Physics and Astronomy, University of St. Andrews KY16 9SS, UK}
\date{Received 22/06/2012 - Accepted 28/08/2012}


\abstract {Most of the mass in protoplanetary disks is in the form of gas. The study of the gas and its diagnostics is of fundamental importance in order to achieve a detailed description of the thermal and chemical structure of the disk. The radiation from the central star (from optical to X-ray wavelengths) and viscous accretion are the main source of energy and dominates the disk physics and chemistry in its early stages. This is the environment in which the first phases of planet formation will proceed.} 
{We investigate how stellar and disk parameters impact the fine-structure cooling lines \nep, \arp, \on, \cp\,and H$_2$O rotational lines in the disk. These lines are potentially powerful diagnostics of the disk structure and their modelling permits a thorough interpretation of the observations carried out with instrumental facilities such as Spitzer and Herschel.} 
{Following Aresu et al. (2011), we computed a grid of 240 disk models, in which the X-ray luminosity, UV-excess luminosity, minimum dust grain size, dust size distribution power law and surface density distribution power law, are systematically varied. We solve self-consistently for the disk vertical hydrostatic structure in every model and apply detailed line radiative transfer to calculate line fluxes and profiles for a series of well known mid- and far-infrared cooling lines.} {The \on\,63 \mic\,line flux increases with increasing  \luv\,when \lx\,$< 10^{30}$ erg s$^{-1}$, and with increasing X-ray luminosity when \lx\,$> 10^{30}$ erg s$^{-1}$. \cp\,157 \mic\,is mainly driven by \luv\,via C$^+$ production, X-rays affect the line flux to a lesser extent. \nep\,12.8 $\mu$m correlates with X-rays; the line profile emitted from the disk atmosphere shows a double-peaked component, caused by emission in the static disk atmosphere, next to a high velocity double-peaked component, caused by emission in the very inner rim. Water transitions, depending on the disk region they arise from,  show different slopes in the correlation with the \on\,63 \mic\,line.} {}

\maketitle
\section{Introduction}
Protoplanetary disks are the intermediate step between the cloud collapse and the planetary system stage. The understanding of the chemical and physical properties of disks  provides the initial conditions for planet formation. The study of the gaseous component has gained increasing interest in the last few years as it dominates the mass budget in the disk in its early stages and because the increased sensitivity of new instrumentation allowed us to observe it, e.g. Spitzer, Herschel, VLT, ALMA. 

Observations with the Spitzer Space Observatory were used to estimate that the lifetime of the inner ''dusty'' disk is of the order of 10 Myr \citep{Str89,Hai01,Her08}. Finding the corresponding timescale for the gaseous component of the disk is rather complex, because there is not a single gas tracer that can be used; instead recent modelling efforts suggest that a suite of several gas tracers may be necessary to cover the wide range of chemical and excitation conditions in disks \citep{Gor11,Kam11}. \citet{Fed10} looked at accretion signatures to indirectly study the lifetime of the gas component. They found no accretion rate higher than $10^{-11}$ M$_{\odot}$ yr$^{-1}$ for objects older than 10 Myr; assuming an exponential decay they infer a timescale of 2.3 Myr. Mechanisms such as photo-evaporation can remove the gaseous disk on short timescales ($\sim$ 10$^5$ yr$^{-1}$) \citep{Hol93,Hai01,Ale08}. The accretion rates found by \citet{Fed10} would then indicate that the gas and dust lifetime are very similar.

The gas densities and temperatures in disks are suitable for the excitation of IR fine-structure transitions of species such as O, C$^+$, Ne$^+$ and Ar$^+$. Different molecular species (e.g. CO, H$_2$O) probe different disk regions, thereby allowing to build a coherent picture of the whole disk, from a complete suite of observation (see \cite{Ber09} for a comprehensive review). These are among the  diagnostic tools that are used to infer disk and stellar properties from observations with IR satellites like Spitzer and Herschel. The latter covers some major disk cooling lines such as \on\,63 \mic\, and \cp\,157\,\mic.

The origin of \on\,63 \mic\,emission from protoplanetary disks is not unique: it can originate from excitation in the disk and from the interaction of jet/outflows with the circumstellar environment \citep{Cec97,Nis99}. Without spatially resolved observations, the contribution of these two mechanisms is difficult to disentangle \citep{Pod12}.

\on\,63 \mic\,is recognized to be a temperature tracer for the surface layers of protoplanetary disks between 10 and 100 AU \citep{Woi09}. An exploratory study with the DENT grid (\cite{Woi10}, 300000 disk models based on the junction of the thermo-chemical code ProDiMo and the radiative transfer code MCFOST) was carried out by \citet{Pin10} for a sample of T\,Tauri and Herbig stars. They show that FUV radiation, potentially arising from accretion onto the star, plays a dominant role in T\,Tauri stars.  \citet{Mei08} investigated the effect of X-rays, finding a correlation between the line flux and the X-ray luminosity. \citet{Gor08} calculated the \on\,line for models with different FUV and X-ray luminosity, finding a major impact on the line flux by \luv. \citet{Are11} considered X-rays and FUV radiation, finding that for their fiducial solar-type star model X-rays become important above $L_{\rm X} \sim$ 10$^{30}$ erg s$^{-1}$.


\citet{Gla07}, followed by work from \citet{Gor08} and \citet{Erc10}, predicted emission of ionized neon at 12.8 $\mu$m in the atmosphere of an X-ray irradiated protoplanetary disk as a probe of physical conditions of the hot atmosphere ($\sim$ 4000 K) in the inner disk ($<$ 20 AU). The line was detected for the first time the same year with the infra-red spectrometer (IRS) on board of Spitzer \citep{Pas07,Esp07,Lah07}. The origin of the \nep\,12.8 \mic\,line is still under debate. \citet{Gue10} analysed a collection of more than 50 \nep\,detections, and suggested different mechanisms responsible: emission from (a) the disk, (b) the photo-evaporative flow from the disk surface and (c) the jet itself \citep{Shu94,Sha10}. \citet{Sac12} and \citet{Bal12} analysed high spectral resolution \nep\,lines. They infer from the position of the peak with respect to the stellar velocity and from the line profiles that the lines are mostly excited both in photoevaporative flows and in jets. Spectrally resolved observations \citep{Naj09} also found a few sources where the \nep\,is most likely emitted in a disk atmosphere. 

The study of the gaseous disk remains an important challenge, where modelling can provide further insights. Thorough computing of the disk thermal and chemical structure, when coupled to observations of gas tracers of different disk regions, can be used as a powerful baseline to predict the fate of protoplanetary disks. Many protoplanetary disk models include X-rays, as there is general consensus that this radiation plays an important role in the chemistry, and in the thermal balance of the disk. 

In this context several groups developed numerical codes dedicated to the modelling of protoplanetary disk  \citep{Gor04,Gla07,Nom07,Erc08,Woi09,GDH09,Hol09,Woo09}. \citet{Woi10} and \citet{Kam11} investigated the effect of a wide parameter space on disk properties and studied the line emission of species such as CO, O and C$^+$. \citet{Mei08} studied the effect of X-rays on Ne$^+$ and Ne$^{2+}$, different ionization stages of S, C, C$^+$ as well as O for a low mass star disk model. \citet{GDH09} investigated the role of FUV (6 $<$ h$\nu <$ 13.6 eV), EUV (13.6 eV $<$ h$\nu <$ 100 eV) and X-rays (h$\nu >$ 100 eV) on the emission of [ArII] at 7 $\mu$m, [NeII] at 12.8 $\mu$m, [SI] at 25 $\mu$m, [FeII] at 26 $\mu$m, [OI] 63 $\mu$m and also pure rotational lines of H$_2$ and CO. They also investigated  the properties of EUV/X-ray driven photo-evaporative flows in the upper layers of protoplanetary disks and their potential impact on diagnostic lines.

\citet{Fog11} studied the effects of grain settling and Ly$\alpha$ line scattering on the chemistry finding that both effects are important for molecules like CO, CN, HCN and H$_2$O, and in general for the carbon and oxygen molecular chemistry. \citet{Hei11} showed that viscous heating and turbulent mixing also impact the molecular layer, thereby modifying the thermo-chemical conditions.

The scope of this paper is to follow up earlier work of \citet{Are11} and to study the relative importance of X-rays and FUV in setting the thermal and chemical conditions in the disk. This work will investigate how selected atomic and molecular diagnostics respond to the different energy input. Meijerink et al. (2012), hereafter Paper\,I, show the respective roles of X-rays and FUV  stellar radiation on the disk physical and chemical structure. The grid consists of 240 models, where we varied \lx, \luv, dust properties (a$_{\rm min}$, a$_{\rm pow}$) and the power law for the surface density distribution ($\epsilon$). We provide, here in the second paper, an extensive study of the line diagnostics such as \on, \cp, \nep, \arp\,and water and investigate the thermo-chemical conditions behind the excitation of these lines.

This paper is structured as follows: in Sect. 2 we briefly recall the setup of our model grid. In Sect. 3 we list and explain the results obtained, in Sect. 4 we compare our findings to those of previous works. In Sect. 5 we summarize and provide an outlook for future work, in Sect. 6 we draw the conclusions.

\section{The grid of disk models}
We use the thermo-chemical code ProDiMo to compute thermo-chemical disk models in hydrostatic equilibrium. The parameters adopted are listed in Table \ref{par}. We consider a Sun-like T\,Tauri star surrounded by a 10$^{-2}$ M$_{\odot}$ disk. To study the impact of the stellar radiation on the disk physics and chemistry we vary X-ray (\lx) and FUV (\luv) luminosities, dust properties (minimum dust size and dust size distribution power law) and the surface density distribution power law. The dust to gas mass ratio is kept fixed ($\rho$=$\rho_{\rm d}$/$\rho_{\rm g}$=0.01).

Every disk model is composed of 100 radial and vertical points where we solve for the thermal and chemical balance. The resulting vertical temperature profile is then used to compute the vertical disk structure until hydrostatic equilibrium is reached (see Woitke et al. 2009). A self consistent treatment of the disk flaring is  needed to compute reliable optical depths through the disk.

The source of the FUV and X-ray radiation is assumed to be centred at the stellar position (a representative spectrum is shown in paper I). No X-ray scattering was considered. 
We ran the grid on the Millipede cluster of the University of Groningen, see paper I for more information. 

\begin{table}[t]
\centering
\caption{Parameters used in the models.}
\label{par}

\begin{tabular}{l|c|c}
\hline
Quantity & Symbol & Value \\
\hline
\hline
Stellar mass       & $M_*$      & 1 $M_{\odot}$ \\
Effective temperature & $T_{\rm eff}$ & 5770~K \\
Stellar luminosity & $L_*$      & 1 $L_{\odot}$ \\
Disk mass          & $M_{\rm disk}$ & 0.01 $M_{\odot}$ \\
X-ray luminosity (0.1-50 keV)& $L_{\rm X}$   & $0,10^{29},10^{30}$\\
&    & $10^{31},10^{32}$~erg/s \\
FUV luminosity & $L_{\rm FUV}$ & $10^{29}$, $10^{30}$,\\
&    &  $10^{31}$, $10^{32}$~erg/s \\
Inner disk radius  & $r_{\rm in}$   & 0.5 AU \\
Outer disk radius  & $r_{\rm out}$   & 500 AU \\
Surface density power law index& $\epsilon$ & 1.0, 1.5 \\
Dust-to-gas mass ratio & $\rho_d / \rho$ & 0.01 \\
Min. dust particle size & $a_{\rm min}$ & 0.1, 0.3, 1.0 $\rm{ \mu m}$ \\
Max. dust particle size & $a_{\rm max}$ & 10 $\rm{ \mu m}$ \\
Dust size distribution & $a_{\rm pow}$ & 2.5, 3.5 \\
power index & & \\
Dust material mass density & $\rho_{\rm gr}$ & $2.5$~g~cm$^{-3}$ \\
Strength of incident ISM FUV & $\chi^{\rm ISM}$ & 1 \\
Cosmic ray ionization rate of H$_2$ & $\zeta_{\rm CR}$ & $5\times 10^{-17}$~s$^{-1}$ \\
Abundance of PAHs relative & $f_{\rm PAH}$ & 0.12 \\
 to ISM & & \\
viscosity parameter & $\alpha$ & 0 \\ 
\hline
\end{tabular}
\end{table}

\subsection{Line radiative transfer}
The line radiative transfer is calculated as explained in \citet{Woi11}, section A.7. We chose an inclination of 45 degrees and a maximum range of $\pm$30 km/s for the width of the lines. The line profile is calculated taking into account thermal and turbulent broadening, the latter is set to $v_{\rm turb}=$0.15 km/s. The flux is calculated for a disk at 140 pc with an inclination ($i$) of 45$^{\circ}$. For a subset of 20 models with varying \lx\,and \luv, and \ami = 0.1, \apo = 3.5 and \eps = 1.5, the line fluxes have been calculated also for $i$ = 0${^\circ}$, 60${^\circ}$, 75${^\circ}$, 90${^\circ}$. The transitions described in this paper are listed in Table \ref{watert}.

\subsection{Data}
The chemical network counts $\sim$1500 reactions, part of it is composed of UMIST reactions \citep{Woo07}, while the reactions involving X-ray primary and secondary ionization of all the elements are taken from \citet{Mei05} and \citet{Ada11}. The treatment of molecular dissociation due to X-ray absorption is explained in Table 1 in \citet{Are11}. Charge exchange reactions between single or double ionized ions with neutral species are taken from \citet{Ada11}. 

In Table \ref{watert} we show the references for the collisional rates used for oxygen, ionized carbon, ionized sulfur and water. The electron excitation cross section for Ar$^+$ and Ar$^{2+}$ were computed in the Iron Project \citep{Hum93}. Collisional rates for Ne$^+$ and Ne$^{2+}$ are taken from the CHIANTI database \citet{Der97,Der09}. \\

\begin{table*}[ht]
\centering
\caption{List of transitions considered in the line radiative transfer. From left to right we list the species name, the transition label, the excitation temperature, the Einstein coefficient for spontaneous emission, the collisional partners included in our model and approximately the region of the disk that the transition traces.}
 \begin{tabular}{l|c|c|c|c|c|c|c}
 \hline
  Species & Transition & $\lambda$ [$\mu$m]& T$_{\rm ex}$[K]  & A (s$^{-1}$) & Coll. partner & Region & Ref.\\
 \hline
 \hline
 O & $^3$P$_1$-$^3$P$_2$ & 63 & 230 & 8.91$\times$10$^{-5}$ & H$_2$, H, H$^+$, e$^-$ & 20-200 AU & 1,2,3,4\\ 
 \hline
 O & $^3$P$_0$-$^3$P$_1$ & 145 & 326 & 1.75$\times$10$^{-5}$ & H$_2$, H, e$^-$ & 20-200 AU & 1,2,3,4 \\ 
 \hline
 C$^+$ & $^2$P$_{1/2}$-$^2$P$_{3/2}$ & 157 & 91 & 2.30$\times$10$^{-6}$ & H$_2$, H, e$^-$ & Outer Disk & 1,2\\ 
 \hline
 Ne$^+$ & $^2$P$_{1/2}$-$^2$P$_{3/2}$ & 12.8 & 1122 & 8.59$\times$10$^{-3}$ & H, e$^-$ & Inner Disk & 5,6\\ 
 \hline
 Ne$^{2+}$ & $^3$P$_{1}$-$^3$P$_{2}$ & 15.5 & 925 & 5.97$\times$10$^{-3}$ & H & Inner Disk & 5,6\\ 
 \hline
 Ar$^{+}$ & $^2$P$_{1/2}$-$^2$P$_{3/2}$ & 6.9 & 2060 & 5.30$\times$10$^{-2}$ & H & Inner Disk & 7\\ 
 \hline
o-H$_2$O & 2$_{12} \rightarrow$ 1$_{01}$  & 179 & 114 & 5.59$\times$10$^{-2}$ & H$_2$, H   & Outer disk & 1,8\\
 \hline
o-H$_2$O & 8$_{18} \rightarrow$ 7$_{07}$  & 63.3 & 1071   & 1.751 & H$_2$, H & Hot belt & 1,8\\
 \hline
o-H$_2$O & 8$_{45} \rightarrow$ 7$_{16}$  & 23.9 & 1615  & 1.05 & H$_2$, H & Inner wall & 1,8\\
 \hline
 \end{tabular}
 \label{watert}
 \caption*{References for the collisional rates: (1) \citet{Sch05}, (2) \citet{Ral09}, (3) \citet{Kre06}, (4) \citep{Sto00}, (5,6) \citet{Der97,Der09}, (7) \citet{Hum93}, (8) The rates for collisions with atomic hydrogen are a scaled version of the H$_2$ data.}
\end{table*}

\begin{figure*}[t!]
 \begin{minipage}[b]{8.0cm}
  \centering
  \includegraphics[scale=0.5,trim=3 0 0 0]{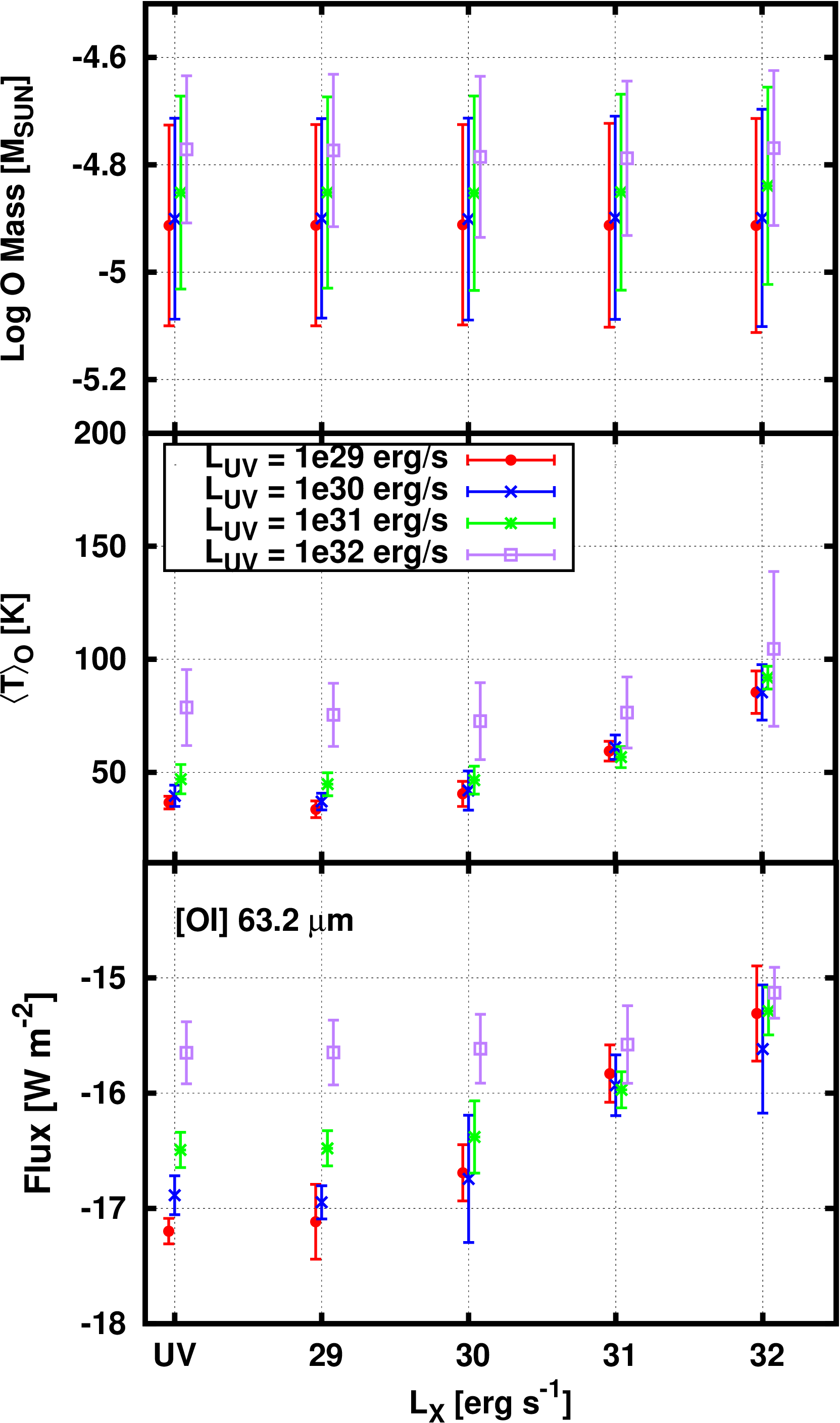}
 \end{minipage}
 \hspace*{8mm}
 \begin{minipage}[b]{8.5cm}
  \centering
  \includegraphics[scale=0.5,trim=3 0 0 0]{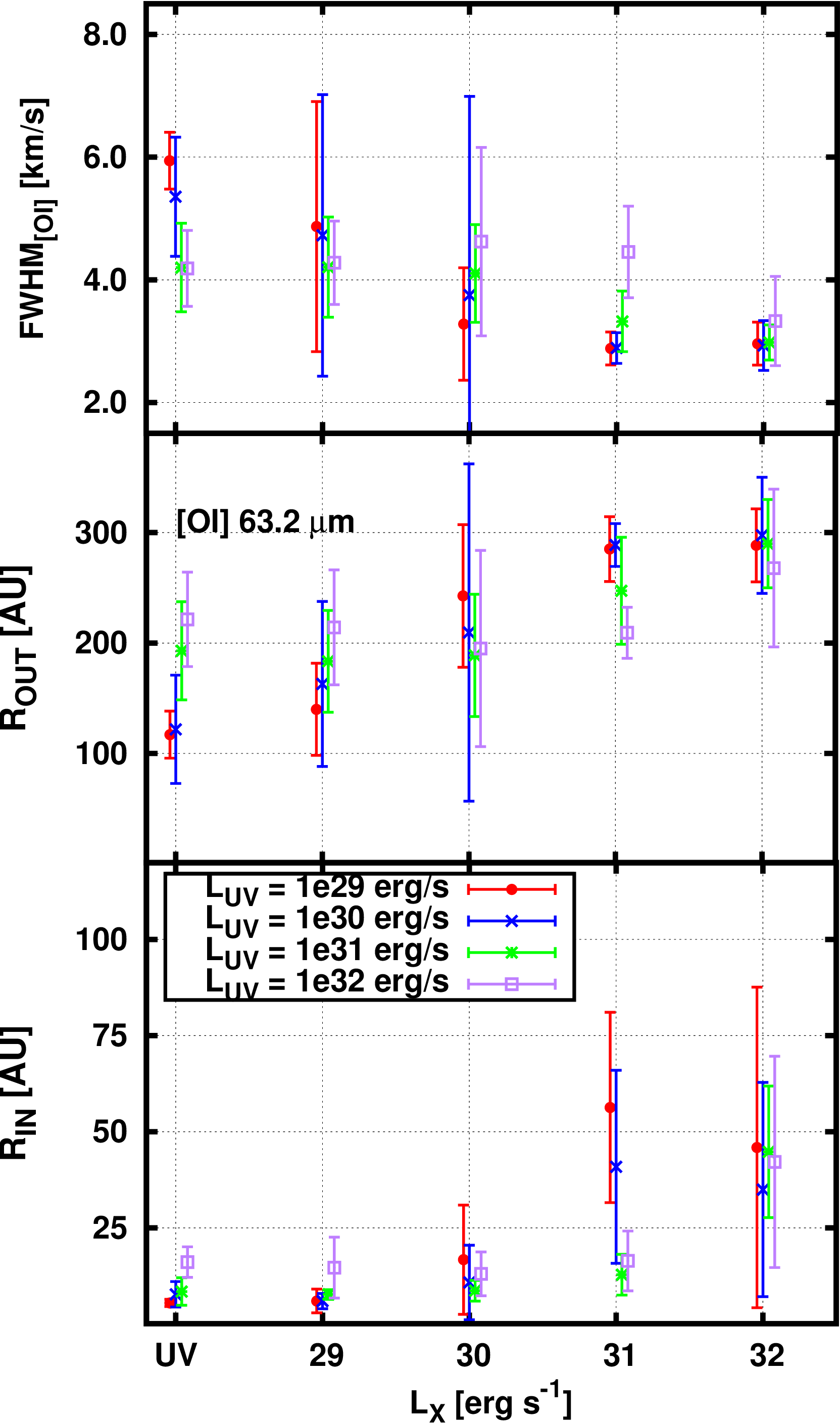}
 \end{minipage}
 \caption{From lower to upper panel on the left: flux of the \on\,63 $\mu$m line, oxygen mass averaged temperature and total oxygen mass in the disk versus X-ray luminosity. See Sect. 3 for the description of the error bars. Since the line is optically thick, its flux is sensitive to the thermal conditions in the disk. From lower to upper panel on the right: [OI] 63 $\mu$m emitting region inner and outer radius and FWHM of the line. As the emitting region gets pushed to larger radii, the FWHM of the line decreases accordingly.}
 \label{oxy}
\end{figure*}
\section{Results}\label{res}
 In this section we describe the resulting line fluxes for \on\,63 \mic , \cp\,157 \mic, \nep\,12.8 \mic\,and \arp\,7 \mic. For a better understanding of the thermal and chemical conditions under which these species emit, we also give a description of the mass averaged gas temperature (defined below) and the total species mass. In addition, ProDiMo computes both the location of the line  emitting region (defined below) and the FWHM of the line. These quantities will be used to investigate how the different stellar and disk parameters affect the emission of the coolants.

The dominant factors controlling these lines are \luv\, and \lx. Hence the quantities we will describe in the following are averaged among models with fixed \luv\, and \lx. The effect of varying \ami, \apo\,and \eps\,is then shown in form of an error bar. Defining a ''series'' as a suite of models with fixed \lx\,and \luv\,but different \ami, \apo and \eps, we count 12 models in each series. This way, we consider for a given physical quantity \emph{Q} in the series k, the simple arithmetic mean $\overline{Q}_k$. To avoid overshoot in the error bars due to a handful of models (n $<$ 10) that did not reach optimal global convergence, we make use of the weighted mean for plotting purposes. The squared distance of $q_{k,i}$ from the arithmetic mean is then calculated and the inverse is used as weight ($\omega_{k,i}=1/\sigma_{k,i}^2$) to compute the weighted average of the quantity Q for the k-th series:
\begin{equation}
Q_W = \frac{\sum_{i=1}^{12} q_{k,i}\cdot\omega_{k,i}}{\sum_{i=1}^{12} \omega_{k,i}}
\end{equation}
We computed the variance $\sigma_{k}^2$ for each series and the error bars in the plots represent then the $2\sigma$ deviation from the $Q_W$. This approach is used for all the quantities discussed in this paper (flux, mass  averaged temperature, mass of species sp, $R_{\rm in}$ etc.)

We also use the mass averaged temperature for a species sp, defined as:
\begin{equation}\label{mat}
 \tave_{\rm sp} = \frac{\int T_{\rm gas}(r,z)\,n_{\rm sp}(r,z)m_{\rm sp}dV}{\int n_{\rm sp}(r,z)m_{\rm sp}dV}
\end{equation}

We calculate the mass averaged temperature only in the emitting region of the considered species: applying a 1D escape probability line radiative transfer, we calculate the radial (\rin\,and \rou) and vertical coordinates that enclose the region where up to half of the total flux of the line is emitted. This volume is then used to perform the integration in Eq. \ref{mat}. This allows to relate the temperature  directly to the flux of the emitted line.

\subsection{Oxygen fine-structure emission at 63 \mic}
\begin{figure*}[th!]
 
  \centering
  \includegraphics[scale=0.35]{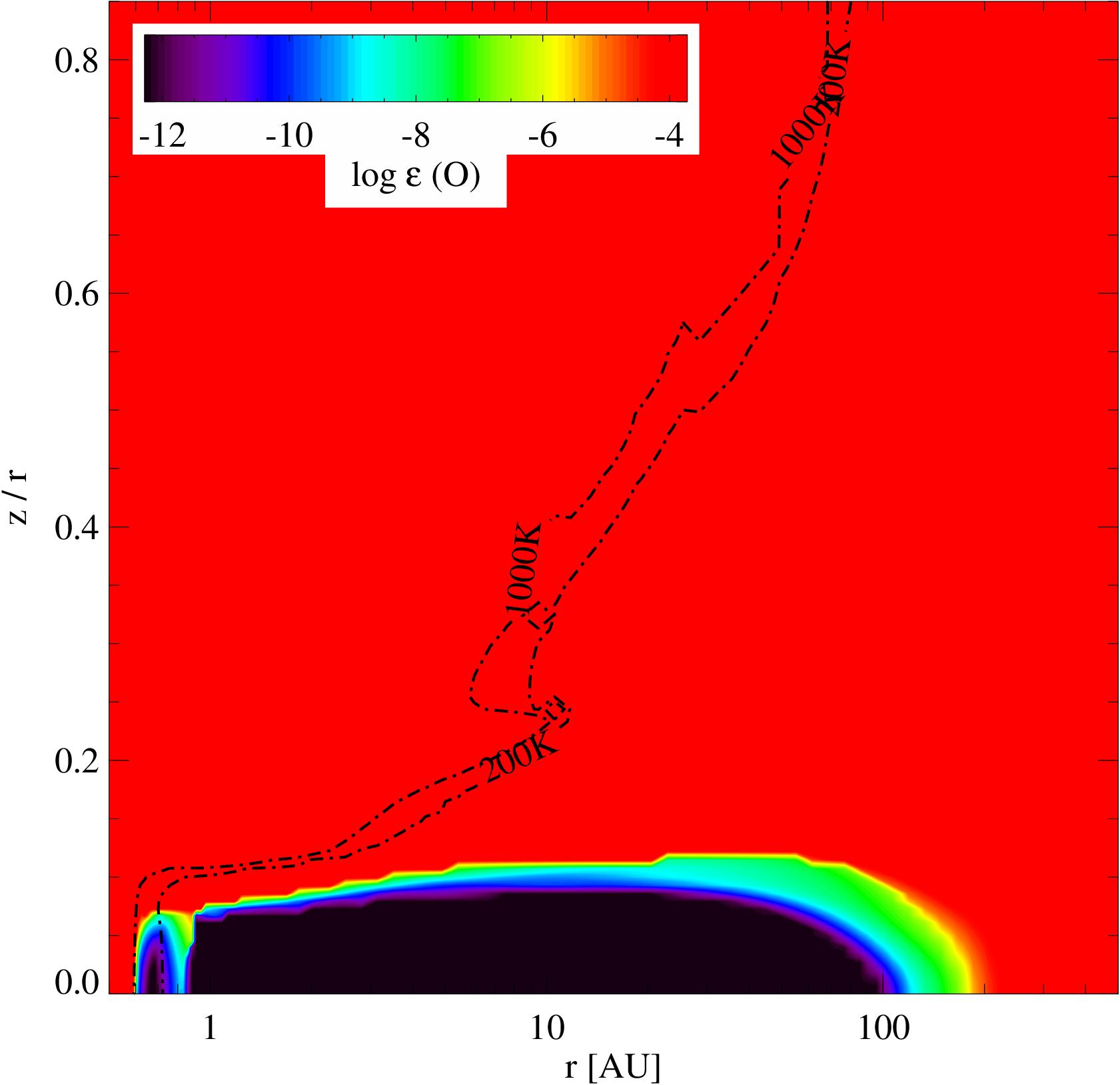}
  \includegraphics[scale=0.35,trim=-3 0 0 0]{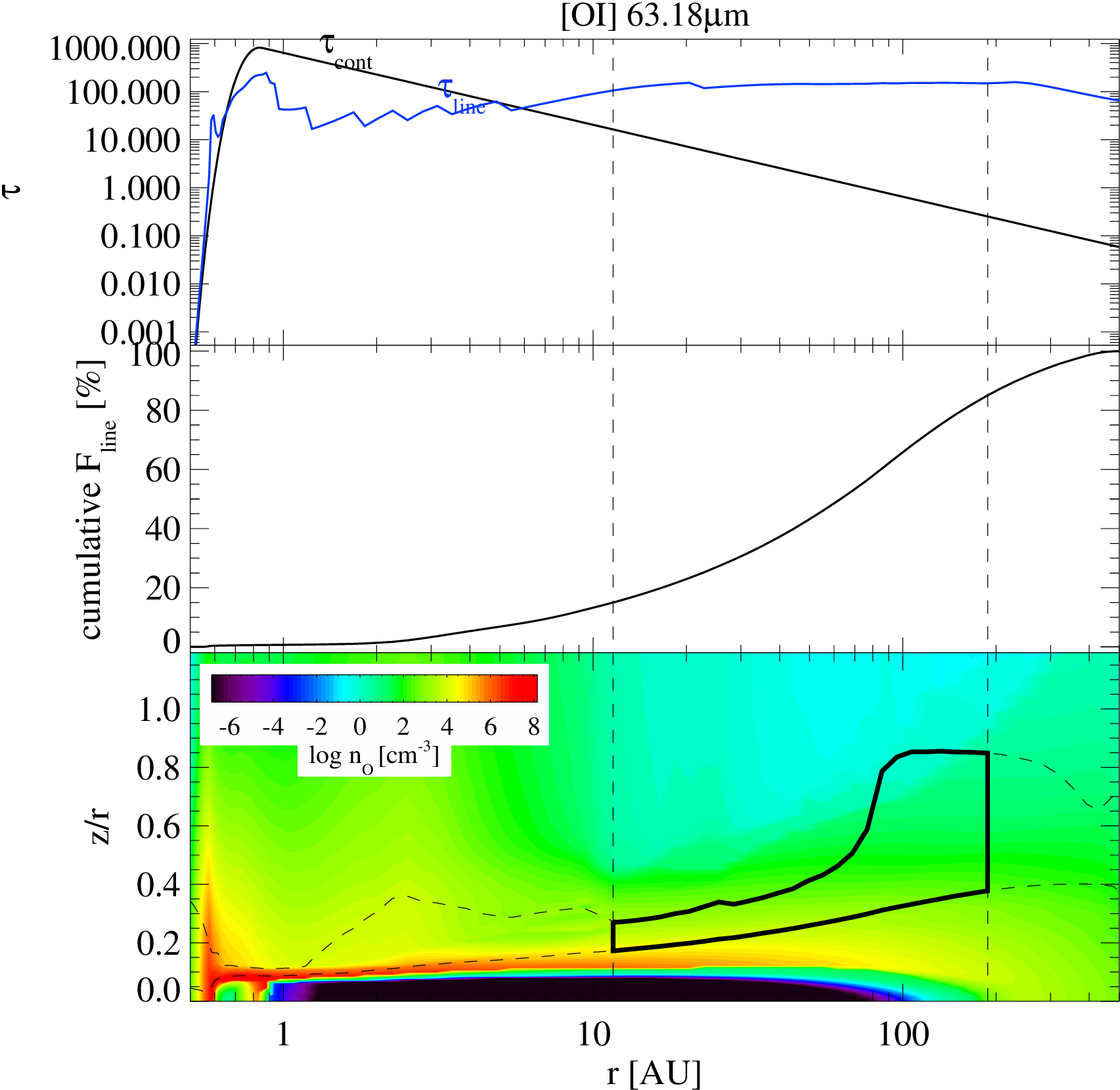}%
  \includegraphics[scale=0.28]{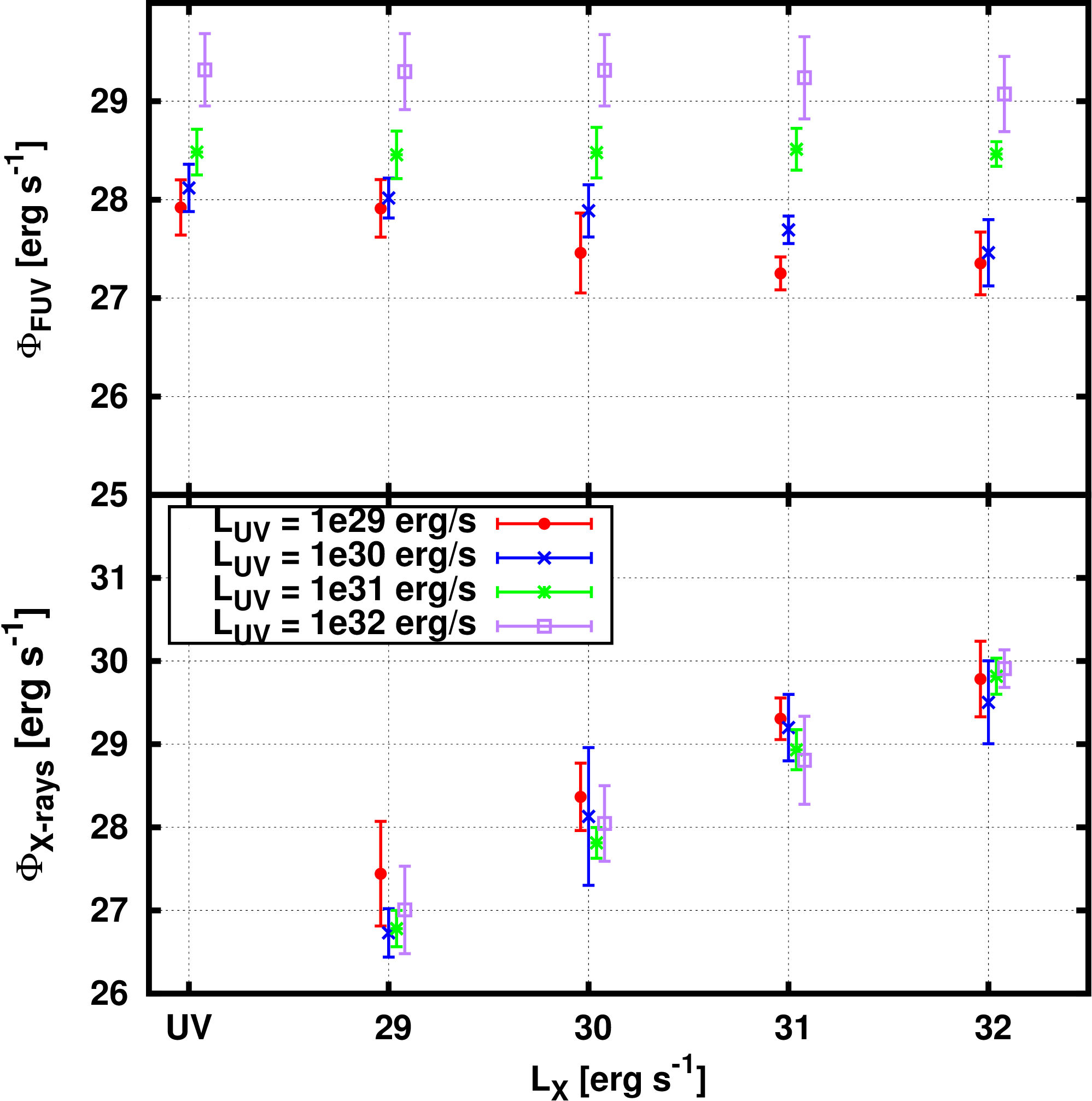}
  \caption{Left panel: O abundance (relative to hydrogen nuclei). The black contours are iso-temperature curves at 200  and 1000 K. Central panel: optical depth of the line (blue) and of the continuum (black) in the top mini-panel, below is the cumulative flux of the line, which shows which percent of the final flux is built and where, the lower mini-panel shows  the oxygen density, the contour indicates where the contribution to the line flux goes vertically from 15$\%$ to 85$\%$ at every radius. Right panel: FUV (top) and X-ray (bottom) heating rates. These rates are obtained integrating the volumetric heating rate $\Gamma$ over the volume of the [OI] emitting region for every model. Averages and error bars have the same meaning as described in section \ref{res}. The first two panels are taken from a representative model with \lx = 10$^{30}$ erg s$^{-1}$, \luv = 10$^{31}$ erg s$^{-1}$, \ami = 0.1 \mic, \apo = 3.5 and \eps = 1.5.}
  \label{OIheat}
\end{figure*}

The lower left panel of Fig. \ref{oxy} shows how the line flux changes as a function of the X-ray luminosity; color coded is the UV luminosity. Plotted is the weighted flux for models with a given X-ray and FUV luminosity and the error bars represent the $2\sigma$ scatter due to the variation of the flux with respect of the other parameters in the grid. 

In the absence of X-rays, \on\,emission is driven by \luv: the 63 $\mu$m line flux increases by a factor $\sim$10 from the lowest FUV model with \luv\,= 10$^{29}$ erg s$^{-1}$ to $L_{\rm FUV} = 10^{31}$ erg s$^{-1}$ and another factor $\sim$ 10 when \luv\, is increased to 10$^{32}$ erg s$^{-1}$. For constant \luv, X-rays always start to impact the \on\,63 $\mu$m emission beyond a threshold of \lx $\sim 10^{30}$ erg s$^{-1}$. The only exception is the highest \luv, where \on\,63 $\mu$m stays constant up to \lx $\sim 10^{31}$. For the highest X-ray luminosity, \on\,reaches a plateau, which represents the maximum value of $\sim 10^{-15}$ W/m$^2$ in our series of models.

The middle left panel of Fig. \ref{oxy} shows the mass averaged temperature of atomic oxygen vs \lx\,in the region of the disk where 50$\%$ of the total flux is emitted. The behaviour is similar to the one discussed for the line flux: the temperature increases with increasing input energy (see also Fig. 2 in paper I). In the UV only models $\left< T\right>_{\rm O}$ slightly increases from T = 35 K in the low \luv\, models, until it reaches about 80 K for \luv $= 10^{32}$ erg s$^{-1}$. The models with the highest FUV luminosity show also the bigger scatter in temperature ($\Delta T \pm$ 20 K). When we let \lx\,increase, the temperature for the low \luv\,models starts to increase when \lx $\sim 10^{30}$ erg s$^{-1}$. In higher \lx\,models, the temperature in the \on\,emitting region is set by X-ray energy deposition, regardless of the \luv\,contribution. 

In the upper left panel of Fig. \ref{oxy}, we show the variation of the total oxygen mass in the disk. The OI mass only varies within a factor 2.5 across all the models. The scatter caused by different dust properties (\ami, \apo) and mass distribution in the disk ($\epsilon$) for a series of 12 models is at most $\sim$ 2.3, hence is higher than the variation of the averaged oxygen mass with \luv in different series of models (maximum variation of a factor 1.4). On the other hand the X-ray luminosity does not affect the total oxygen mass in the disk. This is also explained in section 4.5 of paper I.

The \on\,emission is optically thick in all models. Its flux is then sensitive to temperature variations. 
The mutual FUV/X-ray contribution in creating the thermal conditions that drive \on\,emission becomes clear from Fig. \ref{OIheat} (right panel). We calculated the volume integral for the volumetric heating rate of X-rays and FUV, obtaining the respective heating rates in the oxygen emitting region:

\begin{equation}\label{mat}
 \Phi_{\rm i} = \int \Gamma_{\rm i}(r,z)dV \qquad\rm{[erg\,s^{-1}]}
\end{equation}
Here $\Gamma_{\rm i}$ is the volumetric heating rate in the i-th model for X-rays or FUV radiation in units of erg s$^{-1}$ cm$^{-3}$. We used our previous definition of the oxygen emitting region as integration volume. The FUV emitting rate is the sum of all relevant FUV related processes: photoelectric heating, PAH heating, carbon ionization heating and H$_2$ dissociation heating (see Woitke et al. 2009). In the lower panel, we show how the X-ray heating rate scales roughly proportional to \lx: $\Phi_{\rm X-rays} \sim 0.01$ \lx. In the upper panel, we show the FUV heating rate, we find approximately that $\Phi_{\rm FUV} \sim 0.01-0.001$ \luv. Note how for low \luv\,models increasing X-rays causes progressive decreasing in the FUV heating rate. This is due to a general decrease of the local FUV radiation field at high \lx, which causes FUV heating to decrease, especially PAH heating, which is the main FUV related heating process in the oxygen emitting region (see appendix in \cite{Woi11} for the treatment of PAH heating in ProDiMo). X-ray energy deposition becomes comparable with FUV when L$_X \sim 10^{30}$ erg s$^{-1}$. It is the competition between PAH and Coulomb heating that sets the temperature in the oxygen emitting region, thereby impacting directly on the flux of the line.

The location of the \on\,emitting region varies through the models, especially for the highest values of \lx. Inner and outer radius of the emitting region generally both increase with increasing X-ray luminosity (lower and middle right panel of Fig. \ref{oxy}). Inner and outer radius also increase with \luv. Thereby suggesting a dependence on the total energy input (in the X-rays and FUV band). 

In the top right panel of Fig. \ref{oxy}, we show the full width half maximum (FWHM) of \on\,throughout the model parameter space. The FWHM decreases for increasing \lx, which is in accordance with the changes in the line forming region.


\begin{figure*}[t!]
 \begin{minipage}[b]{8cm}
  \centering
  \includegraphics[scale=0.5,trim=3 0 0 0]{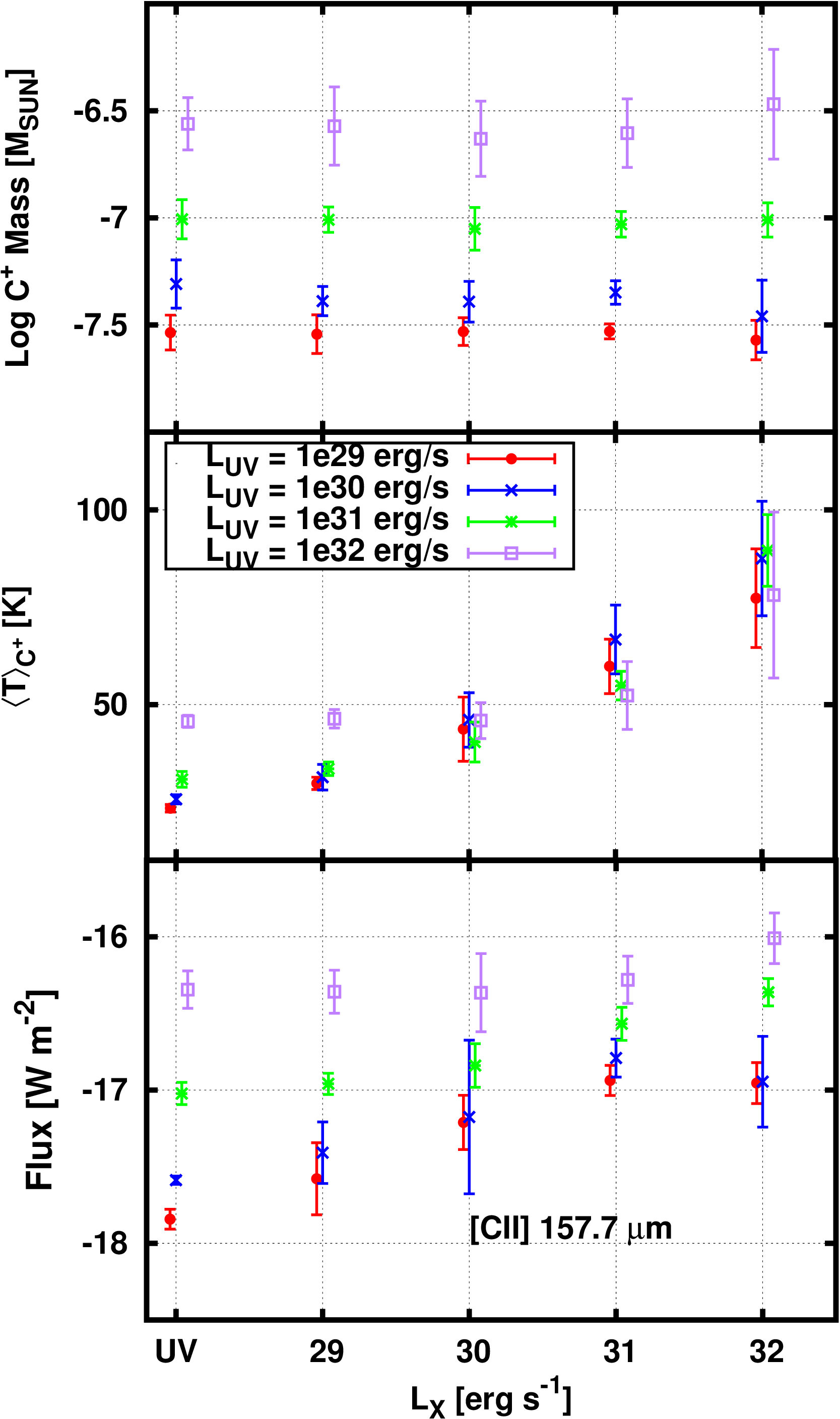}
 \end{minipage}
 \hspace{8mm} 
 \begin{minipage}[b]{8.5cm}
  \centering
  \includegraphics[scale=0.5,trim=3 0 0 0]{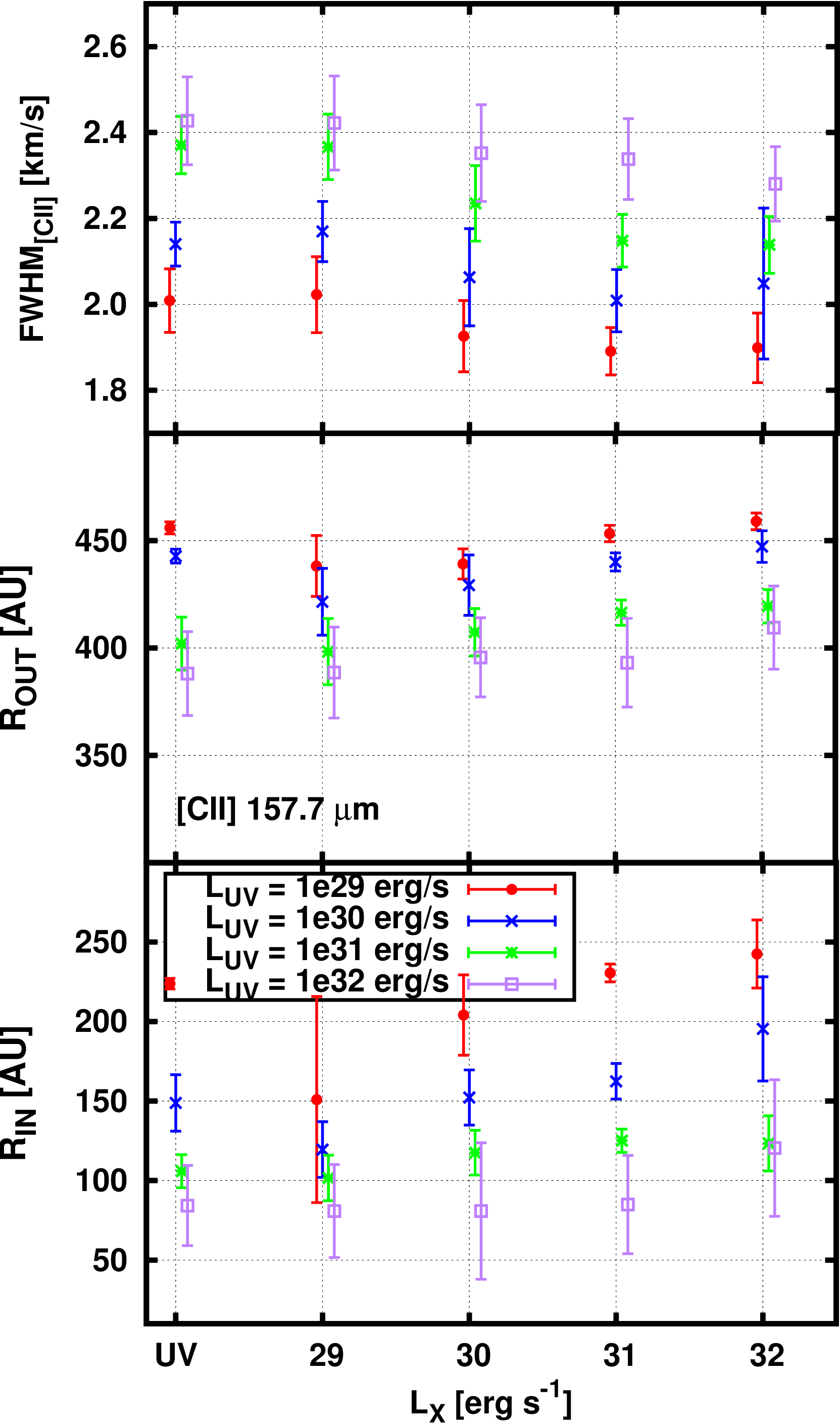}
 \end{minipage}
 \caption{Lower and central panel: inner and outer radius of the emitting region. Top panel: full width half maximum of the line. See Sect. 3 for the description of color coding and error bars. LEFT figure: \on\,at 63 \mic. The oxygen emitting region is generally pushed further as \lx\,increases, this is also reflected by the FWHM, which on average slightly decreases with \lx. RIGHT panel: \cp\,at 157.7 \mic. Ionized carbon traces the outer disk, its emitting region extends inward as \luv\,increases. X-rays do not impact considerably $R_{\rm IN}$, $R_{\rm OUT}$ or the FHWM of the line.}
 \label{carb}
\end{figure*}

\begin{figure*}[t!]
 \begin{minipage}[b]{8.5cm}
  \centering
  \includegraphics[scale=0.5,trim=3 0 0 0]{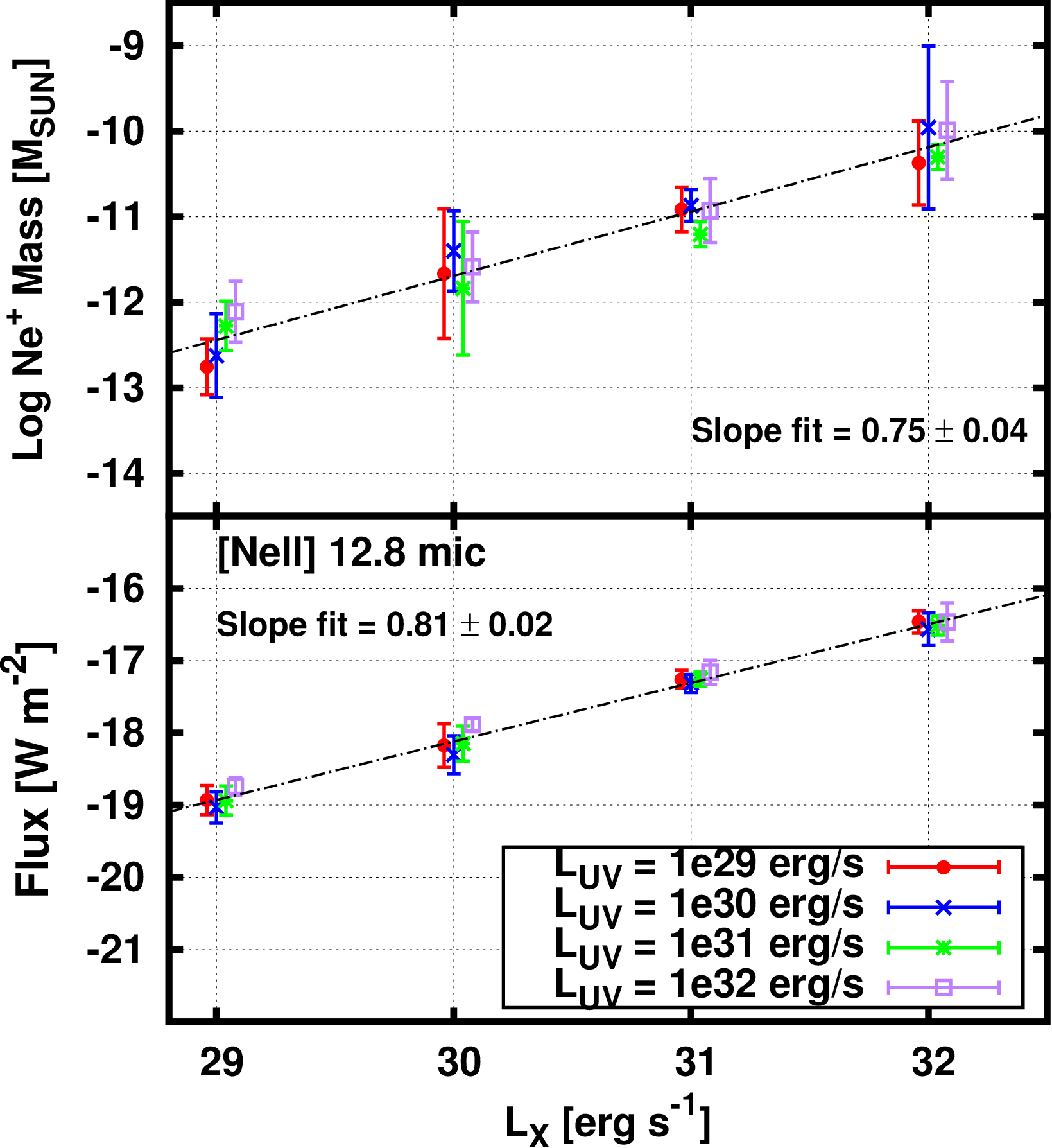}
 \end{minipage}
 \hspace*{5mm}
 \begin{minipage}[b]{8cm}
  \centering
  \includegraphics[scale=0.5,trim=3 0 0 0]{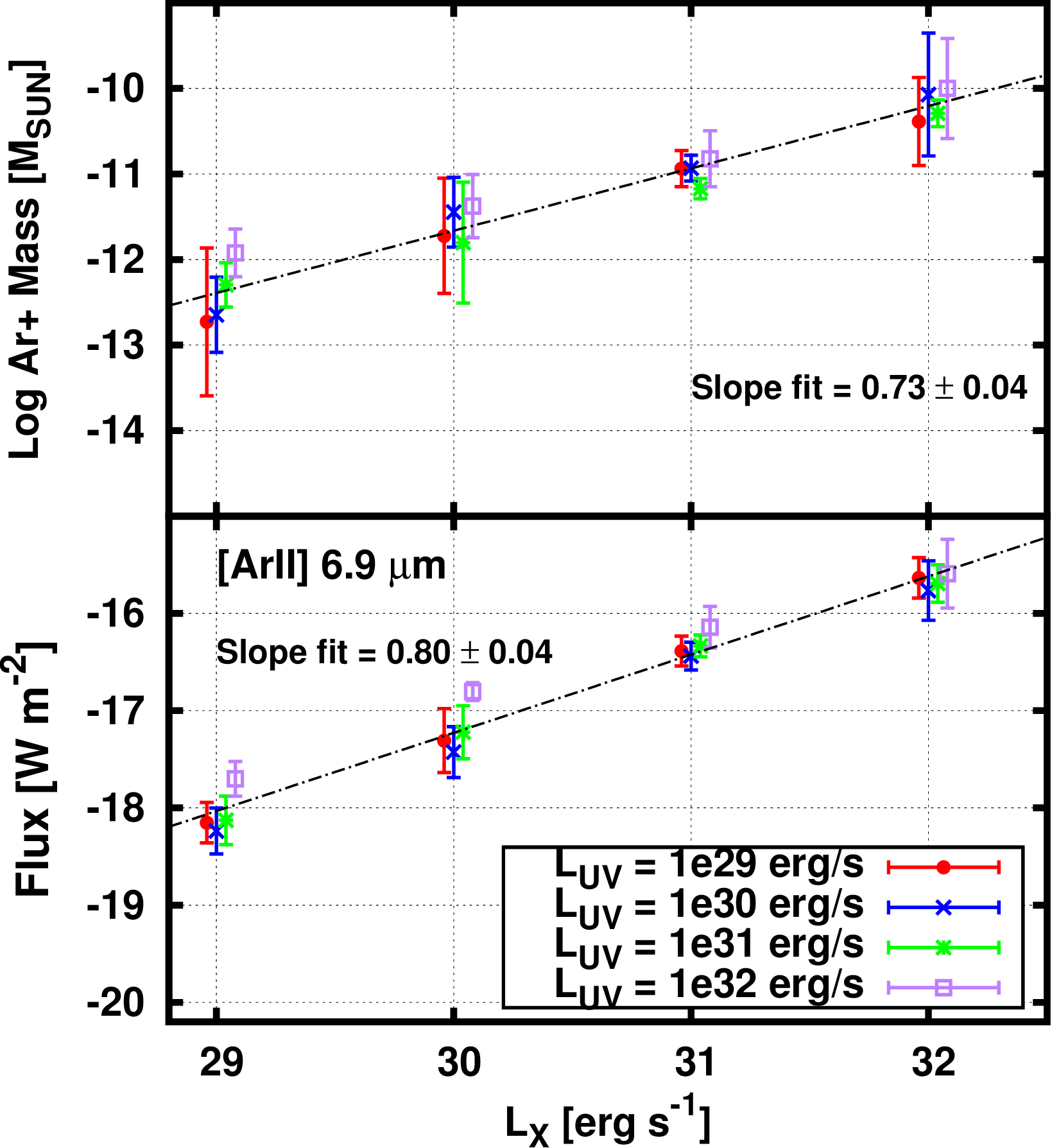}
 \end{minipage}
 \caption{From lower to upper panel: flux of the line, mass averaged temperature and total species mass in the disk versus X-ray luminosity. See section 3 for the description of the error bars. LEFT figure: ionized neon fine-structure line emission at 12.8 \mic. The line is produced in a high temperature X-ray heated environment, where the density is low (n$_{\rm H} \sim 10^{6-7}$ cm$^{-3}$). The line is optically thin and hence very sensitive to the total ionized neon mass, which correlates with \lx. Hence, in our models, \nep\,correlates with \lx. RIGHT figure: ionized argon fine-structure line emission at 6.9 \mic. \arp\, behave very similarly to \nep, and it can be used as a tracer for the same region. }
 \label{nanda}
\end{figure*}

\begin{figure*}[t!]
  \centering
  \includegraphics[scale=0.5,trim=3 0 0 0]{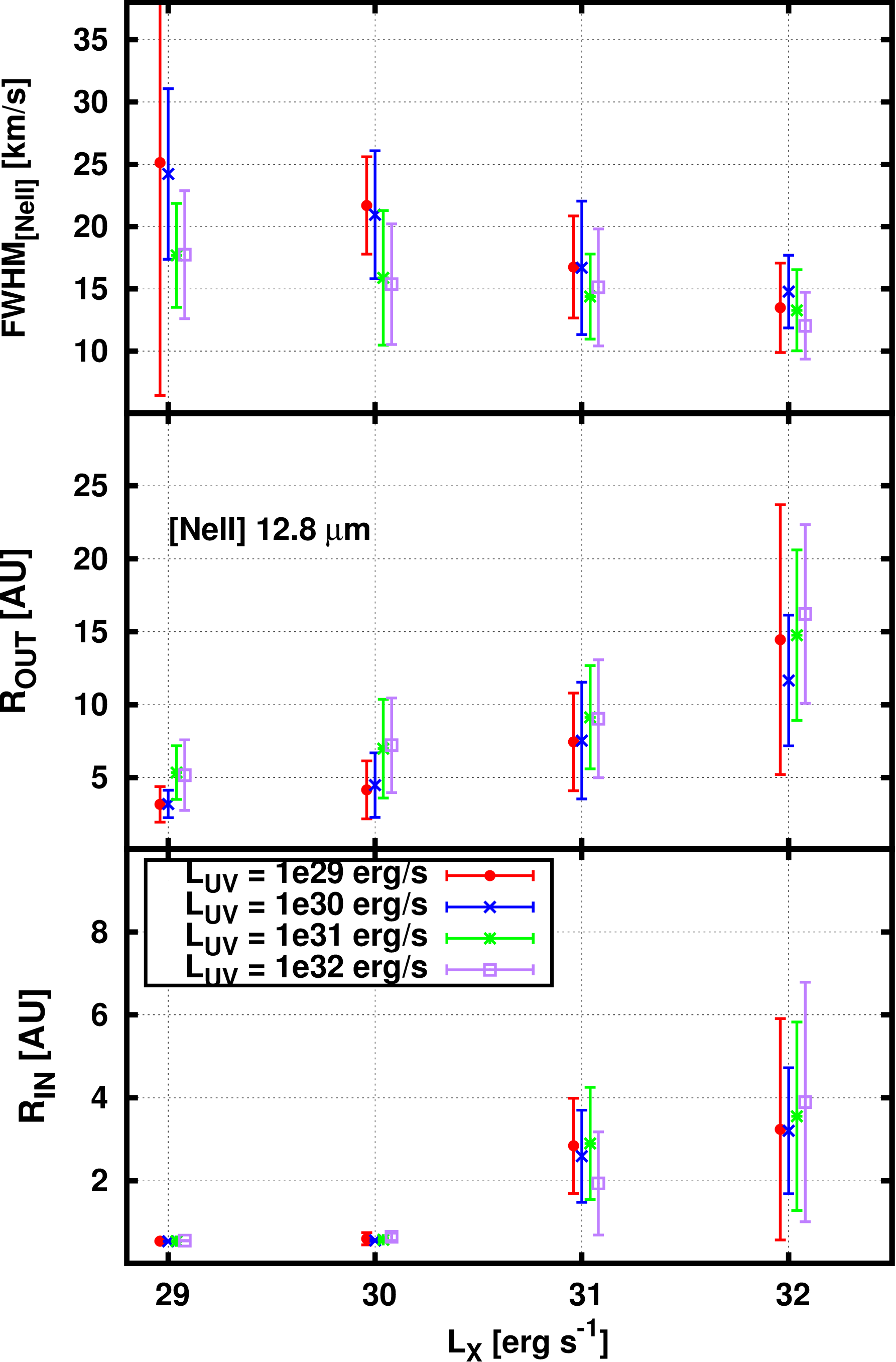}
 \caption{Lower and central panel: inner and outer radius of the Ne$^+$ emitting region. Top panel: full width half maximum of the line. See section 3 for the description of color coding and error bars.}
 \label{nanda2}
\end{figure*}

\subsection{Carbon}
The ionized carbon fine-structure line flux at 157 \mic\,  clearly depends on \luv\, for the UV only models. UV radiation ($\lambda < 1100$ \AA) sets C$^+$ abundances in the outer disk (see paper I). The increasing X-rays can contribute to the line intensity for the lowest two FUV luminosities for \lx\,$< 10^{30}$ erg s$^{-1}$. After this value a ''plateau'' is reached: even though X-rays increase the temperature in the ionized carbon emitting region, the column density in the same region of the disk is decreasing (see the first two panels in the last three rows of Fig. A.24 in paper I), hence the latter effect compensates for the temperature increase. For the high FUV models the column densities do not vary for models with different \lx. Hence the line flux is mainly increasing with \luv, with a minor contribution of \lx, through thermal effects, as in the \on\,63 \mic\,case. 

For \lx $< 10^{30}$ erg s$^{-1}$, the C$^+$ temperature is UV driven and increases for the highest \luv (10$^{30}$ erg s$^{-1}$) by a factor 2 compared to the low \luv\,(10$^{29}$ erg s$^{-1}$). At higher \lx, the C$^+$ temperature is entirely controlled by X-rays driving it up to $\sim$ 100 K in the most extreme case.


In the top left panel of Fig. \ref{carb}, we plot the total mass of C$^+$ as a function of X-ray luminosity. The role of \luv\, in the formation of ionized carbon is clearly shown: the total mass spans roughly one order of magnitude from 3.2$\times$10$^{-8}$ M$_{\odot}$ for the lowest \luv\, to 3.2$\times$10$^{-7}$ M$_{\odot}$ for the highest one. X-rays and the other free parameters affect the total C$^+$ mass to a lesser extent: for a given \luv, m$_{\rm C^+}$  changes at most by a factor 2. 

The \cp\,emission is generally very close to being optically thick. In this regime the line is sensitive both to the column density of the species and to the gas temperature (see paper I, A. 24). Indeed in our models, the ionized carbon flux is mainly dependent on its total mass in the disk. However, if the temperature of the emitting region changes, the flux varies accordingly. These two effects are caused from a combination of FUV radiation, which controls the production of C$^+$ (top left panel of Fig. \ref{carb}), and from X-ray radiation, which, similarly to oxygen, contributes to the thermal balance in the ionized carbon emitting region.

The location of the emitting region is mainly dependent on the FUV flux (Fig. \ref{carb}). In the high luminosity models, C$^+$ emits closer to the central star, starting from $r \sim 80$ AU. For constant \luv, the inner radius of the emitting region \rin, increases only weakly with \lx. However, it shows a strong inverse dependence on \luv:  it increases by a factor 3 from high to low \luv\,models. The outer radius of the emitting region shows the same trend. The different radial location of the ionized carbon emitting region causes the FWHM of the line to change as shown in the upper right panel of Fig. \ref{carb}. The variation is less than a factor 2 from high to low \luv\,models. As expected, the FWHM is not noticeably affected by \lx.

\subsection{Neon}
\begin{figure*}[th!]
 
  \centering
  \includegraphics[scale=0.35]{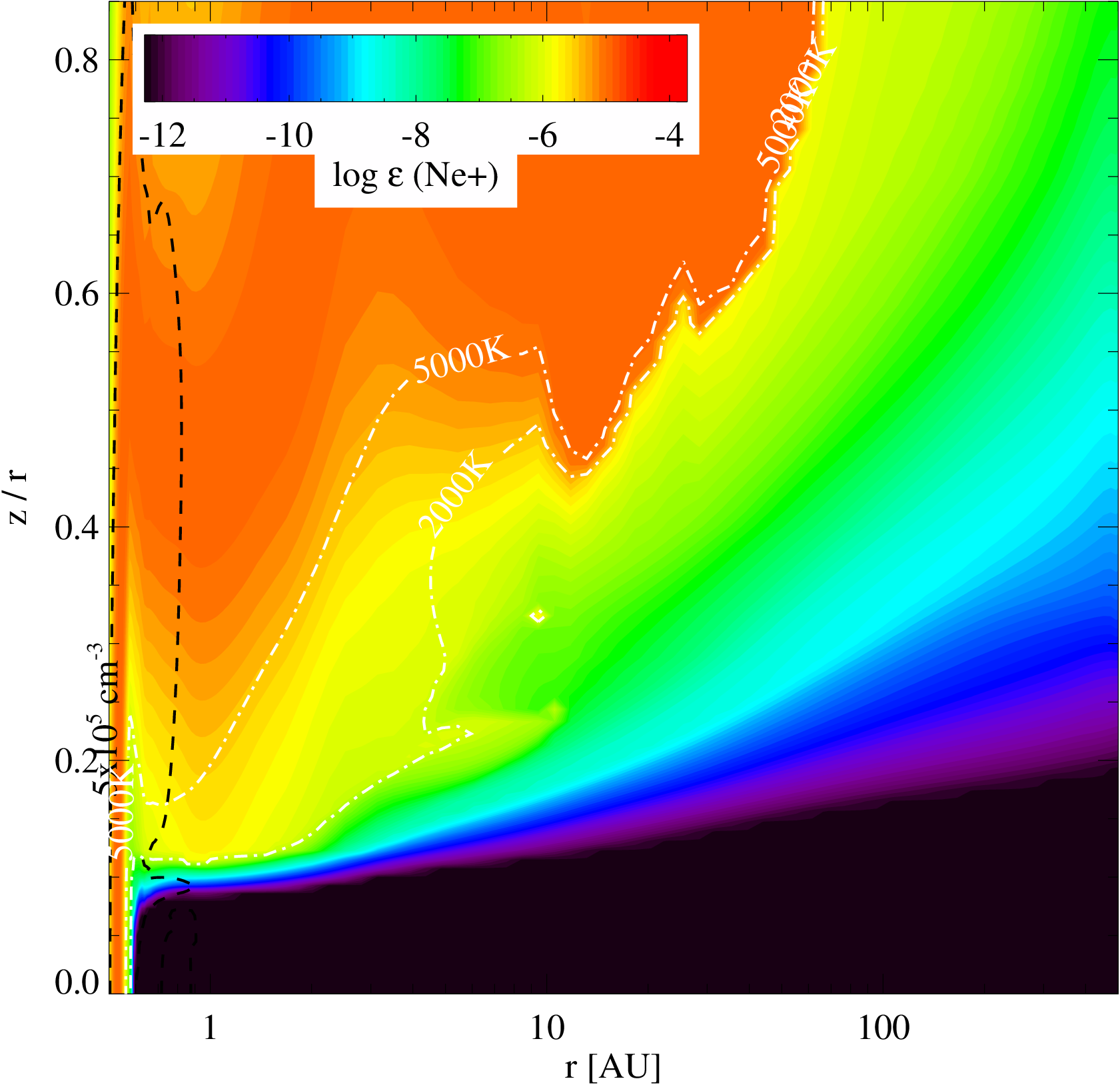}
  \includegraphics[scale=0.35,trim=-3 0 0 0]{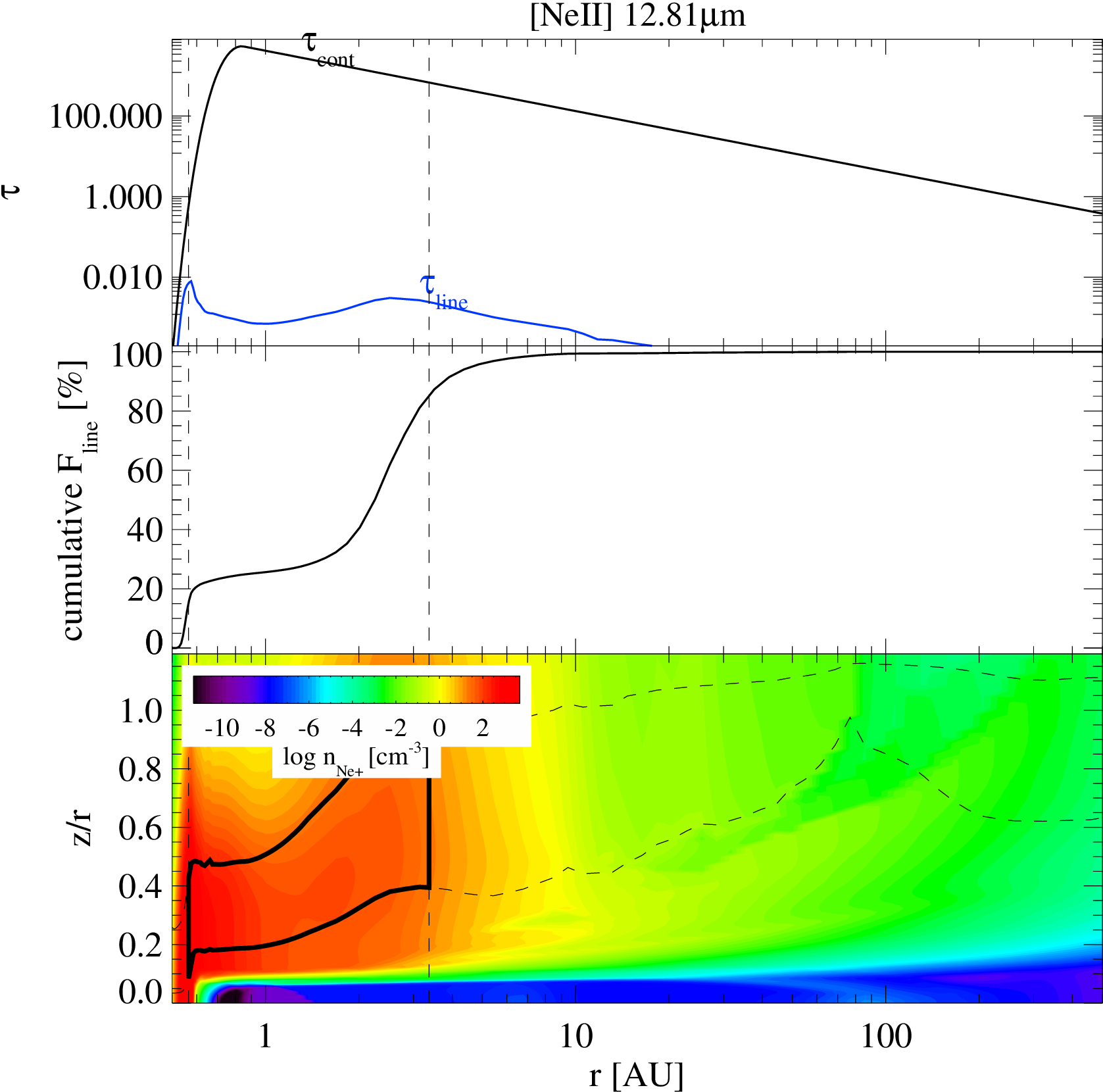}%
  \includegraphics[scale=0.36]{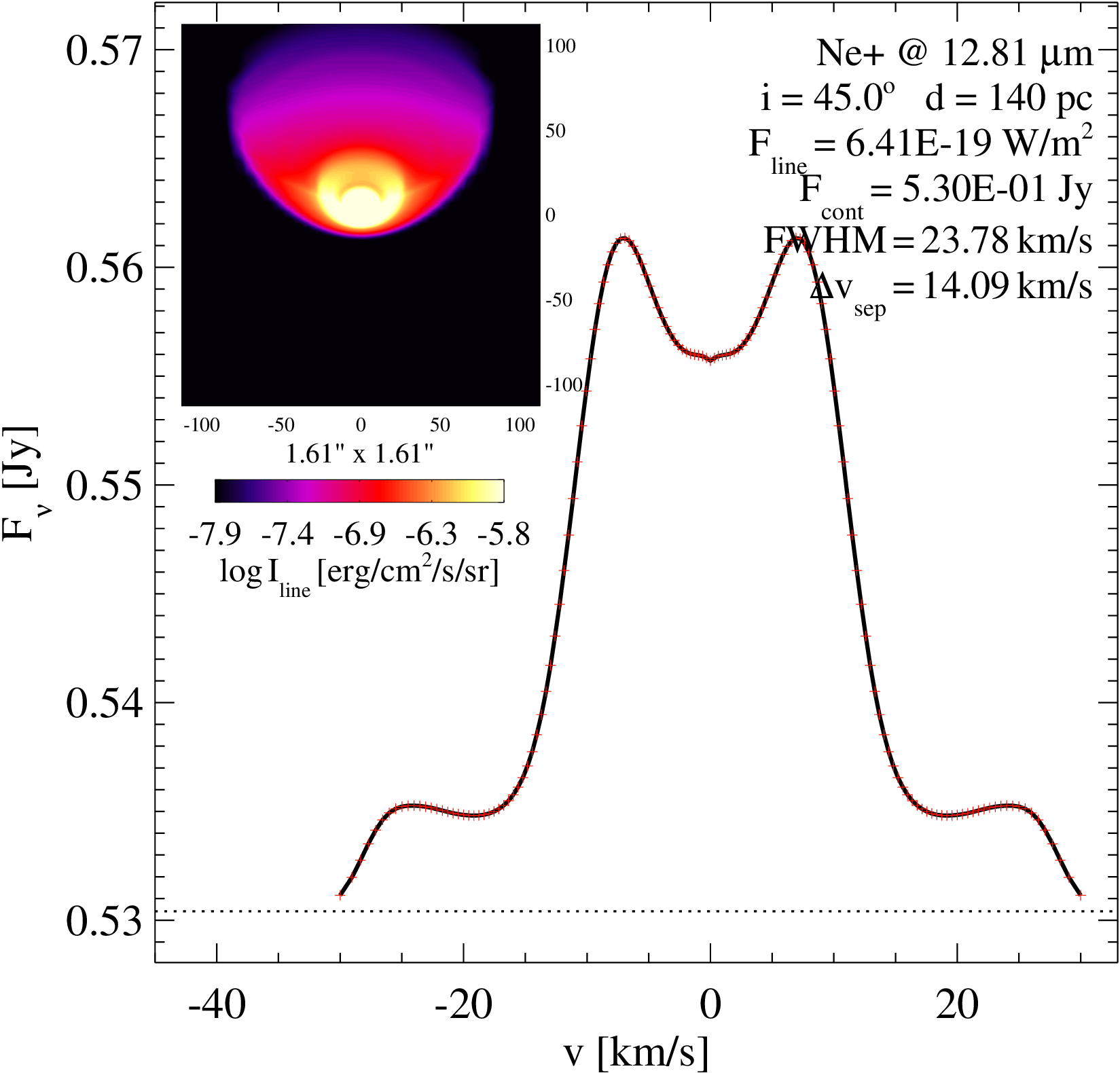}
  \caption{Left panel: Ne$^+$ abundance (relative to hydrogen nuclei). The black contour encloses the region in the disk where the electron density exceeds the critical density for the collisional excitation of the 12.8 \mic\,line. Central panel: optical depth of the line (blue) and of the continuum (black) in the top mini-panel, below is the cumulative flux of the line, which shows which percentage of the final flux is built and where, the lower mini-panel shows again the ionized neon density, the contour indicates where the contribution to the line flux goes vertically from 15$\%$ to 85$\%$ at every radius. Right panel: line profile and image for a distance of 140 pc and an inclination of 45$^\circ$.}
  \label{neline}
\end{figure*}

Ionized neon is expected to be sensitive to the X-ray radiation, because neutral neon has a primary ionization for the K-shell of $\sim$ 900 eV. In the lower left panel of Fig. \ref{nanda}, we show the flux of \nep\,versus \lx. UV only models are not shown as they do not produce significant \nep\, emission. The dependency of \nep\,on parameters other than \lx\,is weak, as reflected in the small error bars for every series of models. The linear fit gives a slope of 0.81 for the correlation of the line flux with \lx.


In the upper panel of Fig. \ref{nanda}, we show the total Ne$^+$ mass in the disk. The higher \lx, the more ionized neon is produced. Other parameters also affect the total Ne$^+$ mass budget, but the overall trend is driven by X-rays. The linear fit gives a slope of 0.75.

The lower and upper left panel of Fig. \ref{oxy} shows clearly that ionized neon fine-structure line emission is controlled by X-ray ionization of neon. The mass averaged temperature, not shown, in the \nep\,emitting region ranges between 2000 K and 5000 K. The electron fractions in the upper layers within 20-30 AU are generally high ($x_{\rm el} \le$ 0.1) but the total electron density where most of the line forms is below $n_{\rm cr}$ (few times 10$^{5} $ cm$^{-3}$) causing the line emission to be mostly not in LTE and optically thin. Its intensity is then regulated by the column density of the species at a given radius, which is clearly set by the X-ray luminosity (see paper I, A. 27). 

The lower and central left panel of Fig. \ref{nanda2} show how \nep\,emission is confined within 20 AU. The emitting region is pushed further out as \lx\,increases. As a result, the FWHM of the line decreases accordingly (upper left panel, fig. \ref{nanda2}) from 25 to 12 km/s.

\subsubsection*{Collisions with H}
Collisions with H are potentially important in the excitation of the fine-structure lines of Ne$^+$. In our grid, we only considered collisional excitation of ionized neon with electrons ($\zeta_{coll,H} = \zeta_1 = 0$). A collisional excitation rate for atomic hydrogen is given in \citet{Bah68}: $\zeta_{coll,H} \sim 2\times 10^{-9}$ cm$^3$ s$^{-1}$ ($\zeta_2$). This value does not take into account the spin interaction between H and the target electron and hence overestimates the rate by almost a factor 10. Following \citet{Mei08}, we therefore adopt $\zeta_{coll,H} \sim 2\times 10^{-10}$ s$^{-1}$ ($\zeta_3$). Using $\zeta_2$ and $\zeta_3$ we run a subset of three models with increasing \lx\, from 10$^{29}$ to 10$^{31}$ erg s$^{-1}$, all other parameters remain unchanged from what refer to as the ''standard model'' (paper I).  
\begin{table}[h]
\centering
\caption{Line fluxes for the 12.8 \mic\,transition of Ne$^+$ considering only collisions with electrons and two different rates for excitation collisions with H.}
 \begin{tabular}{l|c|c|c}
 \hline
  \lx [erg s$^{-1}$] & 10$^{29}$ & 10$^{30}$ & 10$^{31}$ \\
 \hline
 \hline
 $\zeta$ [cm$^3$ s$^{-1}$] &  Flux [W m$^{-2}$] &Flux [W m$^{-2}$] &Flux [W m$^{-2}$]     \\

 \hline
 $2\times 10^{-9}$    & 4.5e-19  & 2.6e-18 & 1.6e-17 \\
 \hline
 $2\times 10^{-10}$    & 2.3e-19  & 1.3e-18 & 7.4e-18 \\
 \hline
 0 & 1.6e-19 & 1.0e-18 & 5.7e-18 \\
 \hline
 \end{tabular}
 \label{hcoll}
\end{table}
Table \ref{hcoll} shows how the collisions with H tend to increase the line flux with respect to the standard models, by a factor $\sim$ 3 for $\zeta_{coll,H}=\zeta_2$ and by a factor 1.3-1.4 for $\zeta_{coll,H}=\zeta_3$.

Collisions with hydrogen affect the total \nep\,flux by at most a factor of a few, and the trend is systematic through all models. Given the uncertainties in the collisional rates, we did not further consider excitation by H collisions in the grid of disk models presented here. 
\subsection{Argon}
In Fig. \ref{nanda} (right figure, lower panel), we show the fluxes for ionized argon fine-structure emission at 6.9 \mic. In the same figure, central and upper panel show the mass averaged temperature and the total Ar$^+$ mass in the disk. Except for the absolute value of the line flux  ($F_{\rm Ar^+}$ $\sim 10\times$ $F_{\rm Ne^+}$), all the results described above for Ne$^+$ apply. The thermal properties in the \arp\,emitting region are the same as described above for Ne$^+$. The total mass of Ar$^+$ follows the same behaviour, in terms of magnitude and dependence from \lx, as found for Ne$^{+}$.

\begin{figure}[t!]
 \begin{minipage}[b]{3.5cm}
  \centering
  \includegraphics[scale=0.5]{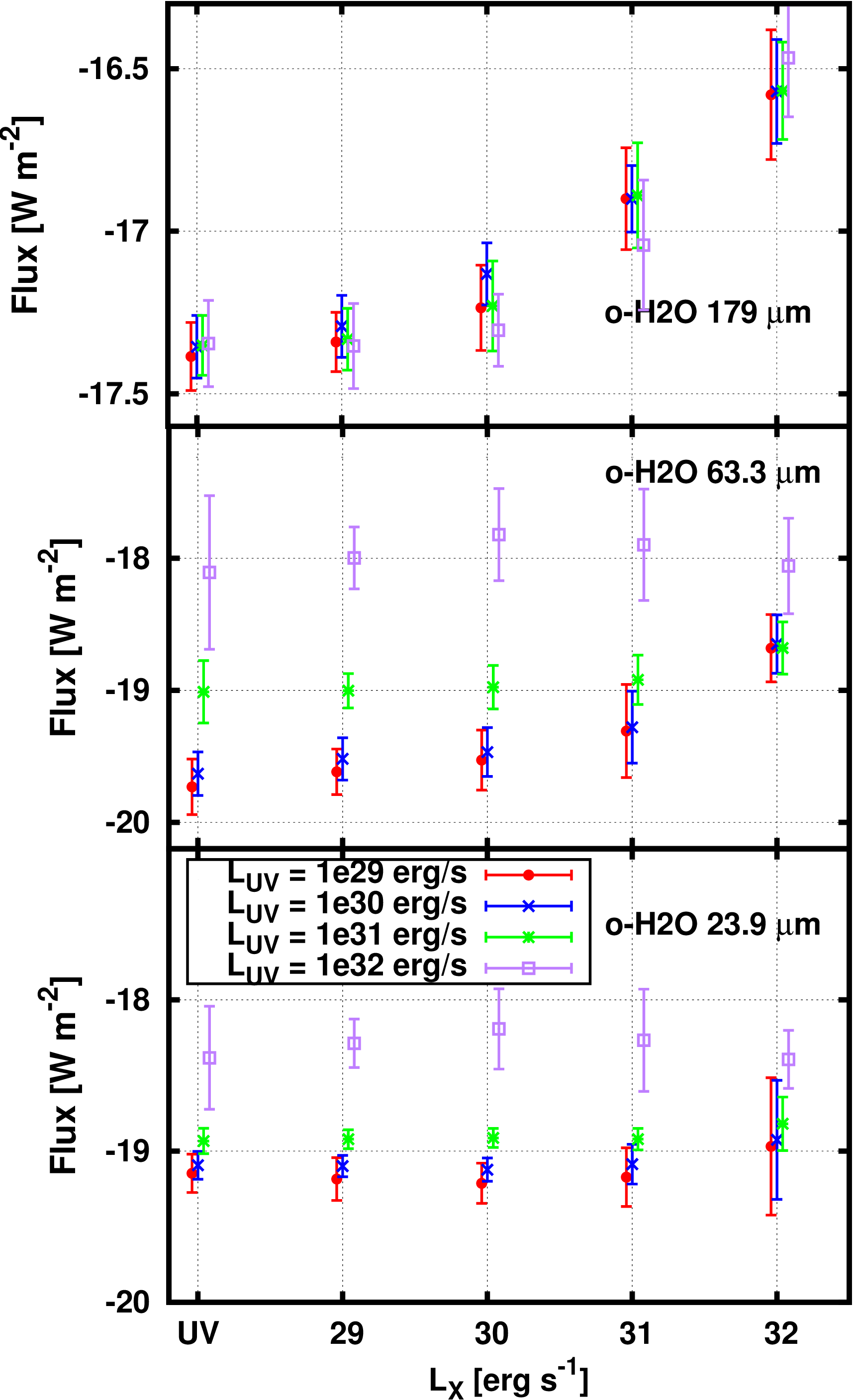}
 \end{minipage}
 \caption{From lower to upper panel: water line fluxes for the 23.9 \mic, 63.3 \mic\,and 179 \mic transition versus the X-ray luminosity. Colour coding is the same as shown in Fig. \ref{oxy}. The X-ray impact is more evident as we move toward the outer disk.}
 \label{watvsx}
\end{figure}

\subsection{Water}
We included in the line radiative transfer also a few rotational water transitions (Table \ref{watert}). Given the high number of transitions and their broad range of excitation temperatures, the water molecule emits over a wide range of radii, tracing disk regions that can differ vastly in terms of chemical and temperature properties. Understanding the water excitation conditions offers the chance to extract information on the radial disk structure. Following \citet{Woi09}, we choose to divide the disk into three representative zones: the cold water in the outer disk, the warm to hot water in a zone above the midplane ranging from $R_{\rm in}$ to well beyond the snow line and the warm to hot main gas phase water reservoir behind the inner rim (Fig. \ref{watall}). In Table \ref{watert}, we list the transitions which are well suited to trace these regions: 179 \mic\,(cold region), 63.3 \mic\,(hot surface water) and 23.9 \mic\,(inner wall water).
In Fig. \ref{watvsx}, we show how the different transitions respond to variations in \lx\,and \luv\,in our grid. We can identify three different behaviours. The line emitted in the inner wall (23 \mic\,line, lower panel) is not altered by the X-ray radiation, unless \lx = 10$^{32}$ erg s$^{-1}$, but only responds to \luv. The line formed in the hot water layer shows an oxygen-like behaviour: X-rays increase the line fluxes beyond a given threshold, 10$^{31}$ erg s$^{-1}$ (vs 10$^{30}$ erg s$^{-1}$ for oxygen). The outer disk line flux (179 \mic) also increases with increasing \lx, and it shows no dependency on \luv. To understand these trends, we show in Fig. \ref{watall} the emitting regions for this lines in a subsample of models: starting from the top left panel, \luv\,increases horizontally from 10$^{29}$ to 10$^{32}$ erg s$^{-1}$ while \lx\,increases vertically from 0 to 10$^{32}$ erg s$^{-1}$. The grey scales shows the water density distribution. We overplot the emitting regions for the o-H$_2$O transitions at 179, 63 and 23 \mic. The boxes enclose the disk region where the contribution to the total line flux for a given transition is about 50$\%$. All the lines discussed here are highly optically thick, hence they are sensitive to temperature variations. We also show the \on\,63 \mic\,transition emitting region previously described for comparison. Red contours are iso-temperature curves at $T_{\rm gas}$ = 2000, 200 and 50 K, black contours indicate the $\tau$(1 keV) = 1 and AV = 1 lines. 

\subsection*{o-H$_2$O 23.9 \mic}
This line is emitted in the inner disk in a high density region ($n_{\rm H} \sim 10^{10}$ cm$^{-3}$), where the disk faces directly the stellar radiation and the water abundance relative to the hydrogen atomic nuclei is 10$^{-4}$. The emitting region is optically thick to the X-ray radiation but coincides with the AV=1 line. Increasing \luv\,indeed seems to affect the total emitted flux, while \lx\,radiation does not play a role (Fig. \ref{watvsx}, lower panel). The temperature in the line emitting region is set by FUV heating.
\subsubsection*{o-H$_2$O 63.3 \mic}
The line traces the high temperature regions above the midplane in the inner disk. In this hot water layer, we can recognize a heating pattern similar to that discussed for oxygen in Sect. 3.1 (Fig. \ref{watvsx}, central panel). Only after a given luminosity threshold, 10$^{31}$ erg s$^{-1}$, X-rays are able to contribute to the heating in this region; below \lx = 10$^{31}$ erg s$^{-1}$, FUV heating dominates. The line is formed in the region that lays between $\tau$(1 keV) = 1 and the AV = 1. For X-rays to contribute to the thermal balance in the hot water belt, \lx\,has to be higher than 10$^{31}$ erg s$^{-1}$ and \lx\,$>$\luv.
\subsection*{o-H$_2$O 179 \mic}
This transition is excited in the outer disk ($r \sim$ 30-200 AU). The line flux is more sensitive to the X-ray radiation because water forms in the outer disk through ion-molecule chemistry (see paper I) and X-rays dominate the thermal balance there. The excitation conditions for this transition are then shifted to higher layers with respect to the FUV only models; these are layers where the X-ray optical depth approaches one at 1 keV (and AV $<$ 1) and the temperature is set by Coulomb heating. Fig. \ref{watall} shows how the emitting region ''moves'' upwards as \lx\,increases. Note that apart from the extreme \lx\,model, the 179 \mic\,flux only increases by a factor $\sim$ 3.

\begin{figure*}[t!]
  \centering
  \includegraphics[scale=0.3,angle=-90]{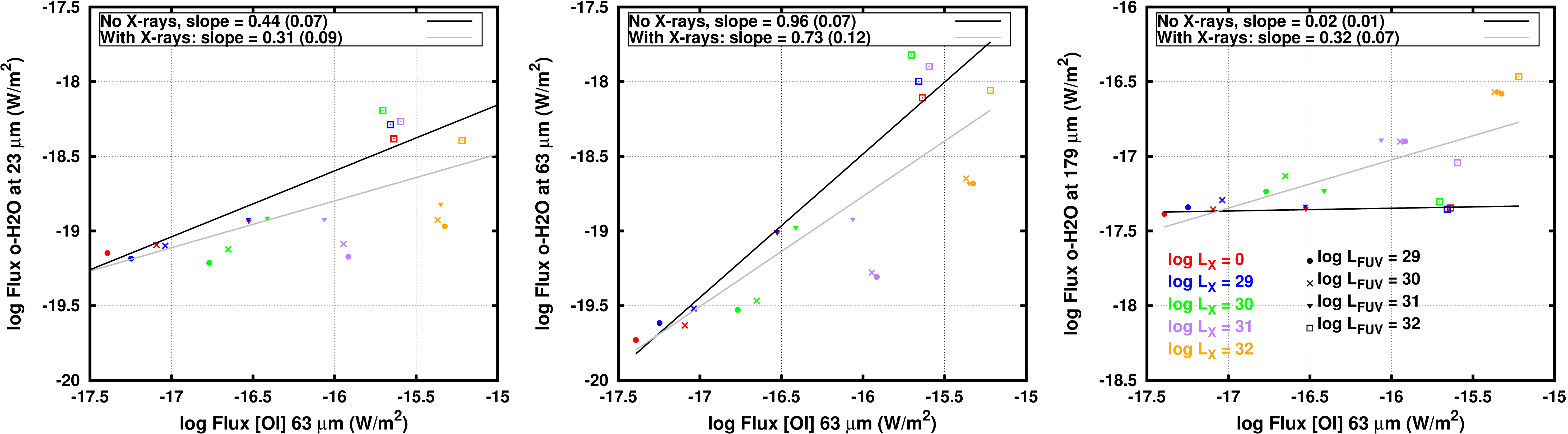}
  \caption{Correlation of water lines with the \on\,63 \mic\, lines. From left to right: 23 \mic, 63 \mic\,and 179 \mic. Color coded is \lx, different symbols indicates different FUV luminosities. The linear fit for the UV only model and the UV+X-rays model are drawn with a black and grey line respectively.}
  \label{wvso}
\end{figure*}

\subsection*{Correlation with \on\,63 \mic.}
In Fig. \ref{wvso}, we show the water lines fluxes versus the \on\,63 \mic\,line flux. At low X-ray luminosities, the 23 \mic \,line weakly correlates with \on\,63 \mic. The correlation is due to the increase of \luv, which heats both emitting regions, causing both line fluxes to increase. The slope of the correlation would be lower than one because, for a given change in \luv, the \on\,flux increases more than the water 23 \mic\,flux, since the latter is shielded by the inner rim. At higher X-ray luminosity, the line flux for \on\,increases, while the o-H$_2$O line does not change anymore. Increasing the FUV luminosity no longer affects the oxygen flux, while the 23 \mic\,water line still increases with \luv. The X-ray models then fall below the previous correlation, causing more scatter in the plot (Fig. \ref{wvso}, left panel). Although we do not provide a ''best fit'', we still show a line for the whole set of models, to illustrate how the slope would change when X-ray models are included.

The same effect can be seen in the correlation between water emission at 63 \mic\,and \on\,63 \mic. Except in this case, the correlation at low X-ray luminosity is almost linear (0.96), because an increasing FUV luminosity causes a uniform increase in both line fluxes. For \lx $\sim 10^{30}$ erg s$^{-1}$, the global correlation is again shallower (0.74).

No correlation is present between o-H$_2$O 179 \mic\,emission and \on\,at low \lx. The former line is indeed not affected by \luv. Though, when \lx\,is higher than $\sim 10^{30}$ erg s$^{-1}$, both line fluxes increase. The slope of the correlation is small and ends up comparable with that of the 23 \mic\,line (0.3), even though the cause is fundamentally different. 

\section{Discussion}
In this section, we discuss our results, comparing them with previous works and analysing their diagnostic potential in the interpretation of disk observations.
 
\subsection{\rm{\on} at 63 \mic}

The \on\,flux is both dependent on \lx\,and \luv. Except in the presence of strong FUV excess (\luv$= 10^{32}$ erg s$^{-1}$) our results suggest that \lx\,has to be larger than 10$^{30}$ erg s$^{-1}$ for the \on\,flux to increase notably. At the threshold, Coulomb heating contributes as much as FUV heating to the thermal balance in the line emitting region.

The total oxygen mass is independent of the X-ray luminosity, while it increases by a factor $\sim$2-3 with increasing \luv\,due to enhanced photodissociation of H$_2$O and OH in the outer disk. Since the line is optically thick, such changes in mass $\sim$ 2-3 do not affect the total line flux. 

The oxygen emitting region is generally pushed further out with increasing \lx, as the disk is overall warmer (PaperI). This is also reflected in the FWHM of the \on\,line, being narrower for high \lx. However, high spectral resolution, such as SOFIA/GREAT is needed (R $\sim$ 75000) to observe this effect. 

\begin{figure}[h]
 \centering
\includegraphics[scale=0.4,angle=-90]{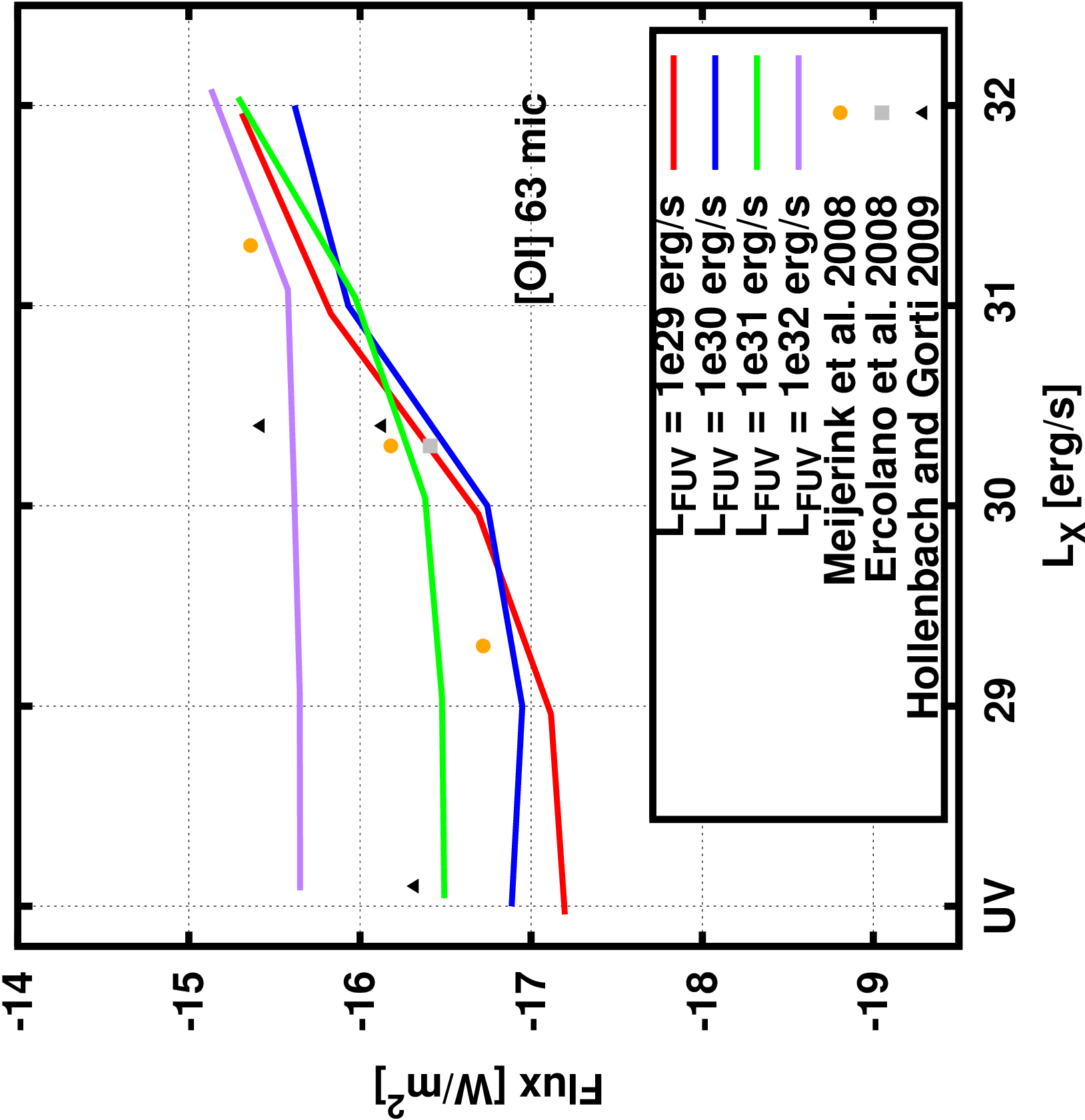}%
 \caption{Our \on\,63 \mic\,results compared with other models.}
 \label{others}
\end{figure}

In Fig. \ref{others}, we show the results from our model grid for \on\,63 \mic\,together with the fluxes found by other authors.
Our results are in good agreement with \citet{Mei08}: their flux at \lx = 2$\times$10$^{30}$ erg s$^{-1}$ matches our values for low \luv\,models. Their prediction for \lx\, = 2$\times$10$^{29}$ erg s$^{-1}$ is lower than ours, but as noted in \citet{Are11} this is due to the absence of FUV in their X-ray irradiated model. They predict a slightly higher flux for their highest X-ray luminosity (\lx\, = 2$\times$10$^{31}$ erg s$^{-1}$). 


The \citet{Erc08} value (log F$_{\rm O}$ $\sim$ -16.41 W/m$^2$ for \lx = 2$\times$10$^{30}$ erg s$^{-1}$ and no FUV) is in very good agreement with our predictions for low \luv. 

Our values are somewhat higher compared with the ones of \citet{Gor08}. In their models, they consider \luv = $5\times 10^{31} \rm erg s^{-1}$ and \lx $\sim 2\times 10^{30} \rm erg s^{-1}$ (fiducial model A). They also calculate a model  without X-rays (model C) and a model D with 10 times higher \luv. They consider different absolute values for \luv, but we reproduce the same \on\,flux variation for models where \luv\,is scaled by a factor 10. Increasing the FUV luminosity by an order of magnitude (from model A to model D) causes the flux to increase by a factor $\sim 5$, which is the same factor we find for such an FUV luminosity variation. The discrepancy in the absolute fluxes might arise from further differences in the model, e.g. their grain size distribution goes from 50 \AA\,to 20 \mic.

The dust parameters we explored in the grid cause a spread in the \on\,flux of less than a factor 3. The same holds for the range of surface density power law exponent explored here (i.e. 1, 1.5). Their impact on the \on\,63 \mic\, line flux is marginal for our grid, but should definitely be taken into account if the goal is to perform multi-wavelength fitting of an individual source. In that case \lx\,and \luv\,are no free parameters, and the secondary disk parameters play the dominant role.

\subsection{\rm{\cp}\,at 157 \mic}

The FUV luminosity controls the line emission, mainly via carbon ionization. Thermal effects impact the line similarly to what happens for \on, but to a lesser extent. The inner and outer radius of the C$^+$ emitting region respond to the FUV radiation.

Our fluxes are higher than the values calculated by \citet{Mei08} and \citet{Erc08}, because these authors do not consider FUV radiation in their models. X-rays alone are not able to sustain enough C$^+$ production for the flux to be on the observable limit of Herschel. However, in many of our models presented here, we overpredict the C$^+$ line: only few sources in Taurus were observed with fluxes  above 10$^{-17}$ W m$^{-2}$ (Howard et al., 2012, in prep.). One reason for this discrepancy might be the shape of the FUV spectrum. As noted already by \citet{Ber03} and \citet{Fog11}, the spectral shape in this range is far from being the simple power law that we adopted for this study. Many emission lines from highly ionized species, especially a very strong Ly$\alpha$ line (10.1 eV, 1215 A), contribute to the total \luv. The lines in a typical T\,Tauri FUV spectrum can carry up to 75\% of the total flux for energies lower than the carbon ionization potential (11.2 eV, $\sim$ 1090 \AA). The C$^+$ mass for a given \luv, depends on how much flux in the FUV range is carried by these lines longward of 1090 \AA. 

In our case of power law continuum, we likely  overestimate, for a given \luv, the continuum flux between 11.2 and 13.6 eV, and this might lead to an overproduction of C$^+$. X-rays do not contribute significantly to the C$^+$ production, because upon X-ray ionization, C$^+$ likely  loses more than one electron through the Auger effect. X-rays also destroy C$^+$ via direct ionization, which has roughly the same rate as neutral carbon X-ray ionization. Subsequent recombination channels form C$^+$, but also neutral carbon. X-rays do not sustain the same level of  carbon ionization as FUV photons (see also the column densities in PaperI).

\subsection{Neon}

\nep\,at 12.8 \mic\,probes the innermost high temperature conditions in the top layers of our protoplanetary disk models. The X-ray photons are  necessary to produce ionized neon and heat the gas there. The temperatures needed to excite the transition via electron collisions ($T_{\rm ex} \sim 1000$ K) can only be achieved in the innermost $\sim$ 50 AU in a X-ray (or EUV) irradiated disk.  

Our models predict a correlation between \nep\,and X-rays; all other parameters explored in this paper contribute marginally to the formation of the line. The slopes in the fits for $M_{\rm Ne^+}$ and $F_{\rm [NeII]}$ vs \lx\,are very similar confirming that the line is optically thin. The higher the X-ray flux, the more extended is the emitting region. This causes the FWHM of the line to anticorrelate with X-rays, changing from $\sim 25$ to $\sim 15$ km s$^{-1}$ when \lx\, goes from 10$^{29}$ to 10$^{31}$ erg s$^{-1}$.
\begin{figure*}[th]
\centering
\includegraphics[scale=0.5]{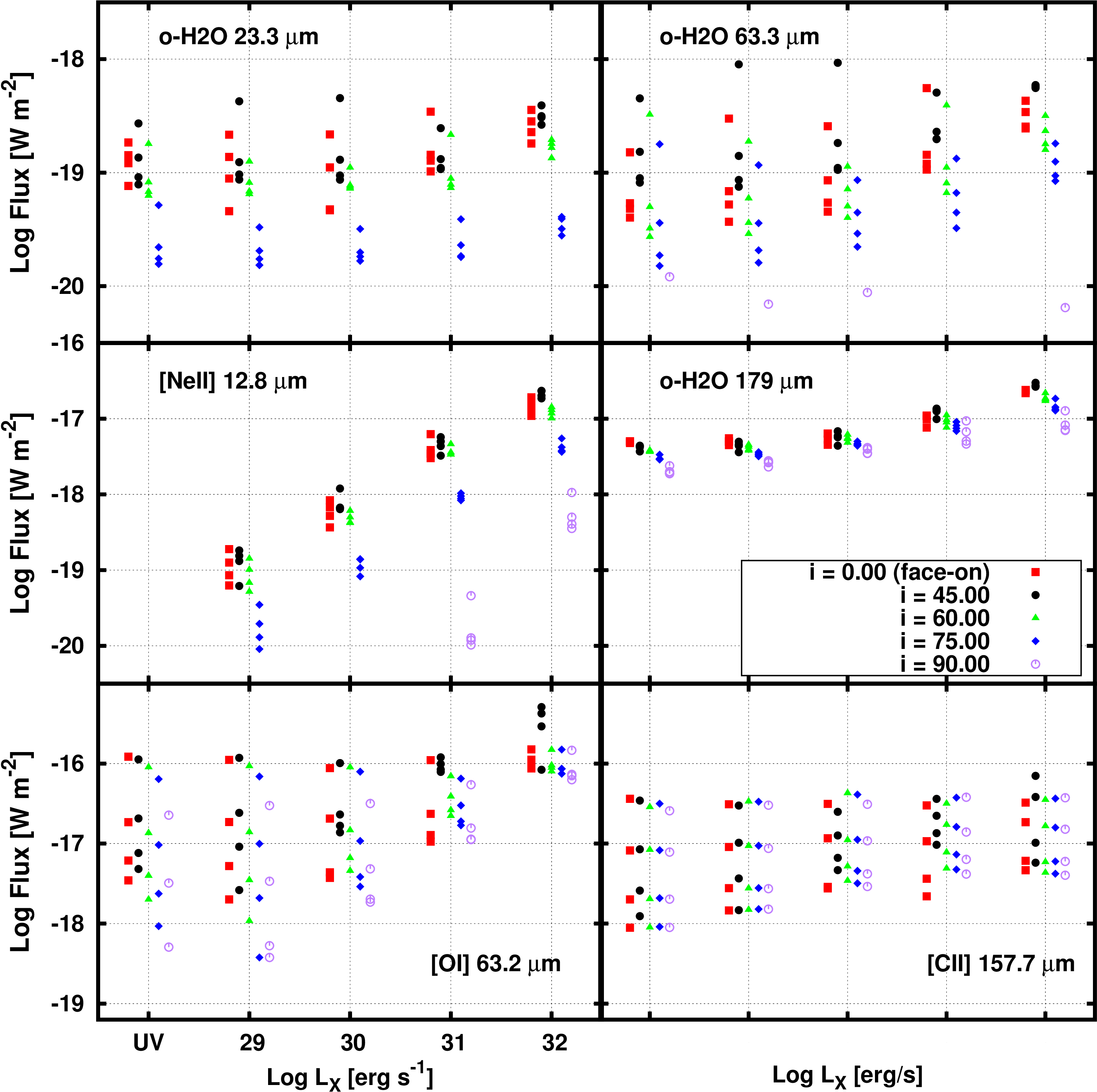}
\caption{Line fluxes for all the tracers discussed in this paper, for different inclinations of the disk: $i$ = 0 (face-on), 45, 60, 75, 90.}
\label{incl}
\end{figure*}

The line is mainly observed in sources that drive outflows or jets \citep{Sac12,Bal12}, theoretical calculations have shown that the line can be indeed excited in such outflows, providing line fluxes and profiles in accordance with the observation \citep{Erc10}.
 
A line probably emitted purely from a disk was observed by \citet{Naj09} toward GM Aur (\lx = 1.2$\times 10^{30}$ erg s$^{-1}$, \citet{Gue10}). The authors note the line is centred at the stellar velocity and has a FWHM of 14 km/s. Our model predictions are consistent with this value (Fig. \ref{nanda2}), and confirm that such FWHM values can be
readily reached. The same line shows a redshifted contribution at $\sim 50$ km/s. Interestingly this could be either a contribution from a jet component (though no known jets/outflow are associated with GM Aur, \cite{Gue10} and references therein) or from the inner rim of the disk (the authors suggest 0.1 to 0.5 AU). The blushifted counter part was not observed because of atmosphere issues and its presence or absence can thus not be assessed. In Fig. \ref{neline}, we show our prediction for a generic model, where the disk inner radius is 0.5 AU. Though we are not able to reproduce the high velocity shift of the second component, we clearly predict the presence of two components in the line profile. The high velocity component originates close to the inner rim (25 km/s correspond to $\sim$ 1-2 AU in our model), while the second lower velocity component originates between ~2 and ~5 AU. 

In this case, the emission is in LTE since the electron density in the inner rim exceeds the critical density for this line. If the high velocity component is confirmed to be symmetric, it could present an interesting diagnostic tool for studying the geometry (height, position, shape) of the inner rim.

In our models the correlation between \nep\,and \lx\,is clear, although we rely on one set of fiducial disk parameters and the X-ray spectrum is scaled to match the required \lx, leaving its shape unvaried. Observations include sources that have different disk properties, which can affect \nep\,as well. This was investigated by \citet{Sch11}. They could reproduce the 1 dex scattering in $L_{\rm [NeII]}$ observed by \citet{Gue10} by varying the hardness of their X-ray spectrum and the flaring angle of the disk model.

\subsection{Ar+}
Our \arp\,6.9 \mic\,results are very similar to the \nep\,results described above. The \arp\,line traces exactly the same region, since the chemistry and excitation of these two species are almost identical. We need to stress here, that the collisional rates used to calculate the \arp\,line fluxes are a factor $\sim$ 10 higher than the collisional rates for Ne$^+$. Given the same basic chemistry for these two species, this is the main reason for the discrepancy in the calculated fluxes. Nevertheless, we are interested in the qualitative behaviour of this tracer, which appears to be an interesting complement to \nep. 

Also \citet{Hol09} give prediction for this tracer concluding that the total fluxes of \nep\,and \arp\,are comparable. 
A study of Spitzer IRS archival spectra was able to detect this line in protoplanetary disks (Szul\'agyi et al., 2012, subm.). More observation are needed to confirm the diagnostic of this tracer. Our estimates for the ionized argon flux are higher than those predicted by other groups \citep{Gor08}. This might be due to the different recipe used for the collisional rates. 

\subsection{Influence of inclination}
So far we restricted the discussion to $i$ = 45$^\circ$, however to estimate the flux variation of the tracers discussed in this paper, we selected a sub-set of 20 models, in which we vary the X-ray and FUV luminosities. The other parameters are fixed: \ami = 0.1, \apo = 1.5 and \eps\,= 1.5. For these 20 models, we run the line radiative transfer for four more different inclinations (0$^\circ$ face-on, 60$^\circ$, 75$^\circ$ and 90$^\circ$). In Fig. \ref{incl} we plot the line fluxes versus \lx, color coded are the different inclinations at which the disk is seen by the observer.

The \on\,flux (Fig. \ref{incl}, bottom left panel) is marginally affected (0.1-0.2 dex) for \lx\,$\leq$ 10$^{30}$ erg s$^{-1}$ and inclinations lower than 75$^\circ$. The flux naturally drops (factor 5-10) for an edge-on disk. When \lx\,$>$ 10$^{30}$ the scatter introduced by models with $i$ different than 45 is $\sim$ 1 dex. Even considering these extreme inclinations, there are no models with \on\,fluxes below 10$^{-17}$ W/m$^2$; hence our threshold for \on is very robust.

The ionized carbon flux and water flux at 179 \mic\,are only marginally affected by different disk inclinations, as it is expected for outer disk tracers. The \nep\,flux (as well as \arp, not shown) do not depend on the inclination when $i <$ 75$^\circ$, while the flux drops by $\sim$ 1 dex for $i$ = 75$^\circ$ and by more than 2 dex for an edge-on disk. Sources with inclination higher than 60$^\circ$ are then likely to contribute to the scatter in the \nep\,line observed by \citet{Gue10} and modelled by \citet{Sch11}.

The water line emission at 23 \mic\,and 63 \mic\,is sensitive to the disk inclination, because the lines are emitted in the inner disk. The scatter is of the order of 0.3 dex or the face-on and $i$ = 60$^\circ$ models, and $\sim$ 1 dex for $i$ = 90$^\circ$. The inclination is then an important parameter to take into account when the correlation between these two water lines and \on\,is evaluated from a random sample of sources.

\subsection{Future}
The results of this grid provide the tool to understand observations of a larger sample of disks. In the context of the GASPS open time key project with Herschel (Gas in Protoplanetary disks, P.I. Bill Dent), observations of \cp\,and \on\,are available for more than 100 sources in Taurus (Dent et al., in preparation).  The \on\,fluxes obtained in the GASPS sample are in qualitative and quantitative good agreement with the predictions made in this work (Howard et al., in preparation). A detailed comparison with the predictions made in this work is the topic of a future paper.

The Spitzer satellite observed the \nep\,line in many sources \citep{Gue10}. Due to the low spectral resolution, it is difficult to assess the location of the emitting region. Recent observations with high spectral resolution instruments  provide a  chance to disentangle the line origin. Currently there are too few sources to carry out a statistical study of the disk emission versus jet emission. Clearly more high spectral resolution ground based observations are needed. 

A range of water lines with different excitation energy probe different disk regions, ranging from hot water in the inner rim to the outermost cold water that originates from photodesorption and ion-molecule chemistry. The correlation
of water lines with the [OI] fine structure line at 63 \mic\, could give information on the disk structure and the excitation conditions in the disk atmosphere. \citet{Riv12}, for example, shows how the correlation between the water and the oxygen line at 63 \mic\,has a slope lower than 1 as predicted here. Dedicated modelling for this and other water observations \citep{Pon10} will also be addressed in a future paper.
 
\section{Conclusions}
\label{conc}
As shown in paper I, changes in the disk structure due to  different luminosity ratios \lx/\luv\,affect the optical thickness and height of the inner rim as well as the position of what we define as the ''second bump''. Hence, models differing only in the X-ray luminosity have different optical depth structure. This modifies the shadowing effects on the outer region of the disk, and impacts directly the thermo-chemical conditions there, hence changing the physical conditions that drive the line emission of different tracers.

Here, we presented an extensive description of fine-structure emission lines of oxygen, ionized carbon, ionized neon, argon and water.
\begin{itemize}
\item The \on\,63 \mic\,line emission is optically thick, and probes thermal conditions of the gas above the molecular layers. The line flux increases with temperature. For \lx\,$< 10^{30}$ erg s$^{-1}$, FUV dominates the thermal balance, above that, \lx $\ge 10^{30}$ erg s$^{-1}$, X-rays dominate.
\\
\item \cp\,157 \mic\,is mainly driven by \luv\,via carbon ionization. X-rays affect the line flux to a lesser extent through Coulomb heating in the \cp\,emitting region. The detailed emission line spectrum and continuum flux levels of the FUV spectrum are important in order to explain observed \cp\,line fluxes.
\\
\item The [NeII] emission from our static modeled disk
atmosphere correlates with X-rays. The line probes thermal conditions in the upper layers of the disk in the region between 1 and 10 AU. Our line profile predictions can be tested with high spectral resolution ground based observation. Especially the presence of a high velocity double-peaked component in the line profile, next to a low velocity double-peaked component, could give informations about the position of the gas inner rim independent of dust observations. This line profile is meant to be compared to sources for which disk emission has been established, as it cannot reproduce line profiles arising from gas in radial motion, such as in photoevaporative flows or jets.
\\
\item \arp\, traces the same disk regions as \nep. Recent work (Szul\`agy et al., 2012, subm.) confirm the detection of this line in protoplanetary disks in few objects.
\\
\item Water emits over a wide range of radii, from the inner to the outer disk. Water line fluxes behave differently depending on the disk region they arise from. The diverse correlations with \on\,63 \mic\,can confirm the validity of the overall chemical and thermal structure of disk atmospheres as modelled with thermo-chemical codes such as ProDiMo. 
\end{itemize}


\noindent
\newline
\tiny \emph{Acknowledgements}. We acknowledge the anonymous referee for his comments and suggestions. We thank Manuel G$\rm{\ddot{u}}$del and Armin Liebhart for their hospitality in Vienna and for very useful discussion on X-rays and Ne$^+$  data and J. Szul\`agy and I. Pascucci for sharing their data on the Ne$^+$/Ar$^+$ line ratio. G. Aresu acknowledges NWO and LKBF for financial support. W.-F. Thi thanks CNES and DIANA for financial support.

\bibliographystyle{aa}
\bibliography{aa}

\appendix

\section{Water}
\begin{figure*}[t!]
 \centering
 \includegraphics[scale=0.9]{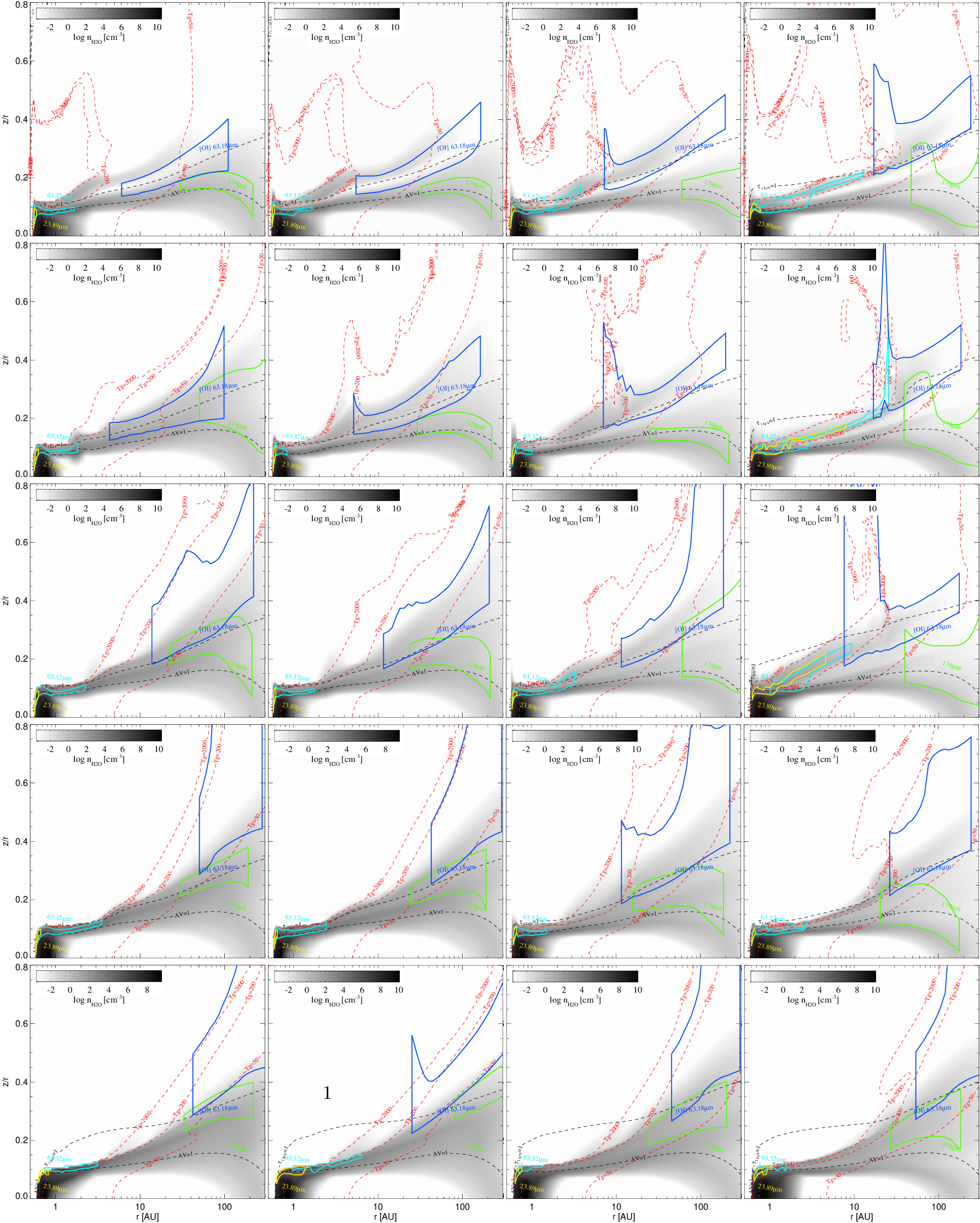}
 \caption{Water emitting regions for a subsample of models with \ami = 0.1 \mic, \apo = 3.5 and $\epsilon$ = 1.5. Yellow, cyan and green mark respectively the emitting region of the o-H$_2$O rotational transitions at 23.8, 63.3 and 179 \mic. Blue is the emitting region of \on\,63 \mic. In grey scale we plot the water density. From left to right are models with increasing \luv\,from 10$^{29}$ to 10$^{32}$ erg s$^{-1}$, from upper to lower panel \lx\,increases from 0 to 10$^{32}$ erg s$^{-1}$.}
 \label{watall}
\end{figure*}

\end{document}